\documentclass{aa}

\usepackage{graphicx}
\usepackage{lineno}
\usepackage{lipsum}
\usepackage{txfonts}
\usepackage[breaklinks=true]{hyperref}
\hypersetup{
  colorlinks   = true,
  urlcolor     = red,
  linkcolor    = blue,
  citecolor    = blue,
  breaklinks   = true
}

\usepackage{subfig}
\usepackage{braket}
\usepackage{float}
\usepackage{multirow}
\usepackage{array}
\usepackage{amssymb}
\usepackage{balance}
\usepackage[toc,page]{appendix}
\usepackage{xcolor}
\usepackage[normalem]{ulem}
\usepackage{changes}

\begin{document}

   \title{Observing the peculiar acceleration of our Solar System with quasar proper motions}

   \author{Phu Huy Nguyen
          \inst{1, 2, 3}
          \and
          Calum Murray \inst{4, 5}
          }

   \institute{Dipartimento di Fisica, Universit\`a di Roma ``Tor Vergata'', via della Ricerca Scientifica 1, I-00133, Roma, Italy
   \and
    Department of Astronomy, University of Belgrade -- Faculty of Mathematics, Studentski trg 16, 11000 Belgrade, Serbia
    \and
    University Observatory, Faculty of Physics, Ludwig-Maximilians-Universit\"at, Scheinerstr. 1, 81679 Munich, Germany
         \and
             Université Paris-Saclay, CEA, IRFU, 91191 Gif-sur-Yvette, France
        \and
             Université Paris Cité, CNRS-IN2P3, APC, 75013 Paris, France
    }

   \date{Received May 18, 2026; accepted August 31, 2026}

  \abstract
{We measured the proper motion dipole of quasars observed by \textit{Gaia} to determine the acceleration of the Solar System with respect to the quasar rest frame. This secular aberration drift provides a purely kinematic measurement of the Galactic acceleration of the Solar System and a mandatory correction for high-precision astrometric reference frames. We characterised the full angular power spectrum of the proper motion field using the pseudo-$C_\ell$ formalism and employed simulation-based inference to jointly constrain the dipole and higher multipole power. Cross-correlation with the \textit{Gaia} scanning strategy, stellar density and stellar proper motion maps was used to diagnose the origin of systematic power beyond the dipole. We applied this framework to both the \textit{Gaia} EDR3 quasar catalogue and the Quaia catalogue. The inferred acceleration is consistent with the previous determination, but the credible intervals widen by factors of 1.5 to 5 when higher-multipole degeneracies are properly marginalised, indicating that previous uncertainty estimates were optimistic. Our best estimate, based on the Quaia catalogue, is ($g_x,\,g_y,\,g_z$) = $\left(1.01^{+0.72}_{-0.72}, \,-4.94^{+0.54}_{-0.55},\,-2.13^{+0.58}_{-0.58}\right) \;\rm \mu as \,yr^{-1}$. This corresponds to an amplitude of $5.55_{-0.53}^{+0.54}\,\rm \;\mu as\,yr^{-1}$, where $\mathbf{g} = \mathbf{a}/c$ is the Solar System acceleration expressed as a proper motion drift in equatorial coordinates. The inferred acceleration is directed towards $(l,\, b) = \left(11.71^{+7.0}_{-6.8},\, -9.01^{+7.0}_{-6.9}\right)^\circ$, equivalently  $(\alpha,\, \delta) = \left(281.39^{+7.8}_{-8.0},\, -22.93^{+6.5}_{-6.2}\right)^\circ$. This is in good agreement with a Galactic centre--dominated acceleration. The acceleration components show no significant dependence on source redshift, providing further evidence of the acceleration's kinematic origin}

   \maketitle
   \nolinenumbers

\section{Introduction}
\label{sec:introduction}

The Solar System orbits the Galactic centre at approximately 230 $\mathrm{km} s^{-1}$ \citep{2019ApJ...885..131R} at a radius of $\rm R \simeq 8.2 \, kpc$ \citep{2019A&A...625L..10G}, which implies a centripetal acceleration ($a$) of $\sim v^2 /\mathrm{R} \approx 2.2 \times 10^{-10} \,\mathrm{m} \,\mathrm{s}^{-2}$ ($\sim 0.7 \,\rm cm/s/yr$) directed towards the Galactic centre. Although this acceleration is very low, it changes the velocity of the observer and hence the aberration of the light received from distant sources, and imprints a predictable dipolar pattern on the apparent proper motions of distant extragalactic sources through the mechanism of secular aberration drift with an expected amplitude of about $4 \rm \; \mu as\,yr^{-1}$ \citep{1995ESASP.379...99B,2003A&A...404..743K}. Quasars are ideal tracers of this effect because quasars lie at cosmological distances and their intrinsic proper motions generated by large-scale structure are expected to be very small, of order $\dfrac{90}{r/1 Mpc} \rm \; \mu as\,yr^{-1}$, as estimated from linear perturbation theory \citep{2019MNRAS.486..145H}. For a quasar at $z \sim 1$, which corresponds to a comoving distance of 3400 Mpc\footnote{This is equal to the comoving distance in a flat universe.}, this yields an intrinsic proper motion of order $10^{-2} \rm \mu as\, yr^{-1}$, some two orders of magnitude below the aberration signal. Therefore, any coherent proper motion pattern observed across the quasar sky encodes the kinematics of the observer rather than the sources.

An accurate measurement of the secular aberration drift is valuable beyond its role as a consistency test. It provides a direct, purely kinematic determination of the Solar System acceleration, and hence of the local Galactic gravitational field, independent of dynamical modelling of stellar tracers. It constitutes a mandatory correction in high-precision astrometry because the third realisation of the International Celestial Reference Frame (ICRF3) applies a galactocentric aberration correction of $5.8\rm \,\mu as\,yr^{-1}$ \citep{2019A&A...630A..93M, 2020A&A...644A.159C}. The reliability of the quoted uncertainty on the acceleration is therefore as important as the measurement itself.

An accurate measurement of the secular aberration drift is valuable beyond its role as a consistency test. It provides a direct, purely kinematic determination of the Solar System acceleration, and hence of the local Galactic gravitational field, independent of dynamical modelling of stellar tracers. It constitutes a mandatory correction in high-precision astrometry because the third realisation of the International Celestial Reference Frame (ICRF3) applies a galactocentric aberration correction of $5.8\rm \,\mu as\,yr^{-1}$ \citep{2019A&A...630A..93M, 2020A&A...644A.159C}. The reliability of the quoted uncertainty on the acceleration is therefore as important as the measurement itself.

The secular aberration drift was first detected with geodetic very long-baseline interferometry (VLBI): \citet{2011A&A...529A..91T} measured a dipole of $6.4 \pm 1.5 \rm \,\mu as\,yr^{-1}$ from the proper motions of 555 radio sources observed between 1990 and 2010, and \citet{2013A&A...559A..95T} refined this to $6.3 \pm 1.1 \rm \,\mu as\,yr^{-1}$ directed within $\sim 7^\circ$ of the Galactic centre. \citet{2017ApJS..233....3T} independently detected the drift at $6.3\sigma$ in the Very Long Baseline Array (VLBA) Extragalactic Proper Motion Catalog of 713 sources. Based on the findings of the International VLBI Service for Geodesy and Astrometry (IVS) Working Group 8 \citep{2019A&A...630A..93M}, a recommended amplitude of $5.8\,\rm \mu as\,yr^{-1}$ was adopted in the construction of ICRF3 \citep{2020A&A...644A.159C}.

The \textit{Gaia} space astrometry mission \citep{2016A&A...595A...1G} enabled the first optical measurement. Its Early Data Release 3 (EDR3; \citealt{2021A&A...649A...1G}) contains astrometric solutions for over 1.5 billion sources, of which approximately 1.6 million have been identified as quasars. The first determination of Solar System acceleration with \textit{Gaia} observations of quasars was carried out by \cite{2021A&A...649A...9G}, who fitted the quasars' proper motion field with a truncated vector spherical harmonic (VSH) expansion and bootstrap resampling to estimate their errors. They obtained an acceleration amplitude of $|\mathbf{g}| = 5.05 \, \pm \, 0.35 \rm \; \mu as \,\mathrm{yr}^{-1}$ directed towards the Galactic centre. This is consistent with the VLBI determinations and with the expectation from the Solar System's Galactic orbit. However, their analysis assumed that bootstrap resampling adequately captures the full uncertainty budget, treating sources as effectively independent. This assumption neglects the spatially correlated systematic errors present in the \textit{Gaia} astrometric solution.

We revisited the measurement of Solar System acceleration from quasar proper motions with two key improvements. First, we employed the angular power spectrum analysis using \texttt{NaMaster} \citep{Alonso_2019} to characterise the full multipole structure of the proper motion field, including cross-correlation with \textit{Gaia} scanning strategy and stellar density and proper motion maps. Second, we adopted a simulation-based inference (SBI) approach in which the dipole is estimated jointly with the higher-multipole power, allowing the uncertainties on the acceleration components to reflect degeneracies with un-modelled large-scale systematics. We applied this framework to both the \textit{Gaia} EDR3 quasar sample and the Quaia catalogue \citep{Storey_Fisher_2024}, a homogeneous quasar sample constructed from \textit{Gaia} and unblurred coadds of the Wide-field Infrared Survey Explorer (unWISE) photometry with explicit modelling of selection effects.

The structure of this paper is as follows. Section~2 reviews the aberration drift formalism. Section~3 describes the datasets we used. Section~4 describes the spherical harmonics decomposition. Section~5 presents the pseudo-C$\ell$ estimator angular power spectrum of the proper motion field, including auto- and cross-correlations with systematic templates. Section~6 describes the SBI framework and its calibration test. Section~7 discusses the implications of our results, and Sect.~8 summarises our conclusions.

\section{Aberration drift} \label{sec:aberration}
Aberration of light refers to the apparent displacement of a celestial object from its true position on the sky, arising from the finite speed of light combined with the motion of the observer. When the observer undergoes acceleration, this displacement varies, causing sources that are intrinsically stationary to exhibit an apparent systematic proper motion. For sufficiently distant extragalactic sources, whose intrinsic proper motions are negligible, this apparent motion encodes exclusively the kinematic state of the observer rather than any physical motion of the sources themselves.  

This apparent proper motion vector induced by the observer's acceleration is given by:
\begin{equation}
    \boldsymbol{\mu} = \mathbf{g} - (\mathbf{g} \cdot \mathbf{u}) \mathbf{u}
    \label{apparent_pm_2}
\end{equation}
where $\mathbf{u}$ is the unit vector pointing from the observer towards the source, $\boldsymbol{\mu}$ is the proper motion vector due to the aberrational drift, or the so-called apparent proper motion, and $\mathbf{g} = \dfrac{\mathbf{a}}{c}$ is the acceleration expressed in units of proper motion such as $\mu$as yr$^{-1}$. The detailed derivation of Eq.~\ref{apparent_pm_2} is discussed in \cite{2021A&A...649A...9G}. Eq.~\ref{apparent_pm_2} is transformed into component form using Cartesian coordinates (x, y, z) within the equatorial (ICRS\footnote{International Celestial Reference Frame. For more information, please visit \url{https://hpiers.obspm.fr/webiers/icrf/icrf.html}}) reference system $(\alpha, \delta)$. The components $\mu_{\alpha*}$ and $\mu_{\delta}$ of the proper motion are obtained by projecting $\mathbf{\mu}$ on the unit vector $\mathbf{e}_{\alpha}$ and $\mathbf{e}_{\delta}$ in the directions of increasing $\alpha$ and $\delta$ at the position of the source as in Fig. 1 in \citet{2012A&A...547A..59M}. The result is 
\begin{equation}
\begin{aligned}
     \mu_{\alpha*} &= -g_x \sin(\alpha) + g_y \cos(\alpha), \\
     \mu_{\delta} &= -g_x \sin(\delta) \cos(\alpha) - g_y \sin(\delta) \sin(\alpha) + g_z \cos(\delta)
\end{aligned}
\label{eq:false_pm}
\end{equation}
where ($g_x$, $g_y$, $g_z$) are the equatorial components of \textbf{g}. 

\section{Data}
\subsection{Gaia} \label{sec:gaia}
\begin{figure}[t]
    \centering
    \includegraphics[width=\linewidth]{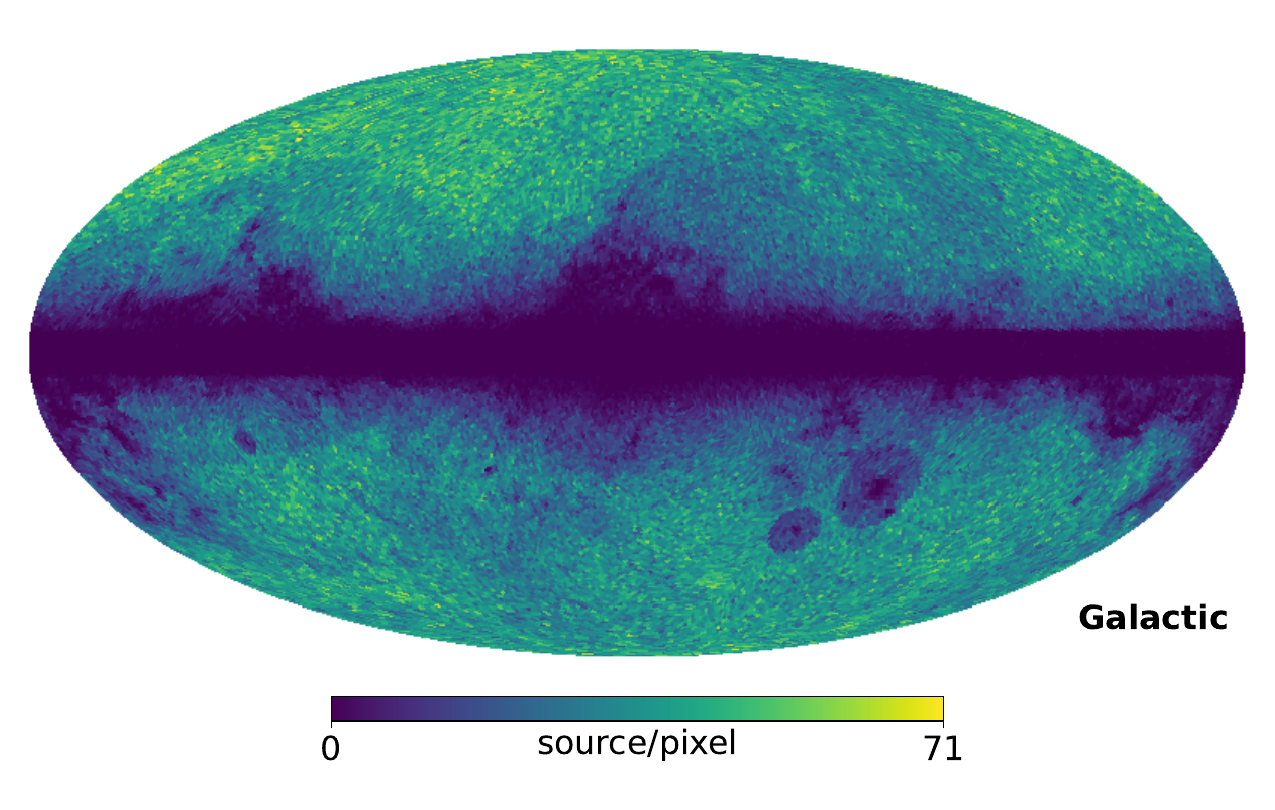}
    \caption{Mollweide map (Galactic coordinates) of the source counts per
HEALPix pixel (\texttt{NSIDE} = 64) for the 1,215,942 Gaia EDR3
quasars with five-parameter astrometric solutions used in this
analysis.}
    \label{fig:nqso}
\end{figure}

For a robust comparison with \citep{2021A&A...649A...9G}, we used the data from \textit{Gaia} EDR3, which is based on 34 months of observations \citep{2016A&A...595A...1G} and provides high-precision astrometry for over 1.5 billion sources, mostly stars in the Milky Way galaxy \citep{2021A&A...649A...1G}. However, in the context of this research, only a quasar subsample or quasar-like sources are required. Sources were selected by cross-matching the full \textit{Gaia} EDR3 catalogue against 17 external quasar and active galactic nucleus catalogues, followed by filtering on astrometric solution quality and requiring that the measured parallaxes and proper motions are statistically consistent with zero, thereby reducing stellar contamination while retaining extragalactic sources.  As a result, 1,614,173 sources that are contained in \textit{Gaia} EDR3 were identified as quasars (Fig.~\ref{fig:nqso}). The full details of the selection procedure are given in \citet{2022A&A...667A.148G}.

The astrometric solutions for individual sources in \textit{Gaia} EDR3 are categorised into three types: two-parameter solutions, five-parameters solution, and six-parameter solutions \citep{2021A&A...649A...2L}.  Since the sources with six-parameter solutions are generally fainter, redder, and have larger systematic errors \citep{2021A&A...649A...5F}, we included only 1,215,942 sources with five-parameter solutions in the analysis. 
\begin{figure}[htbp]
    \centering
    \includegraphics[width=\linewidth]{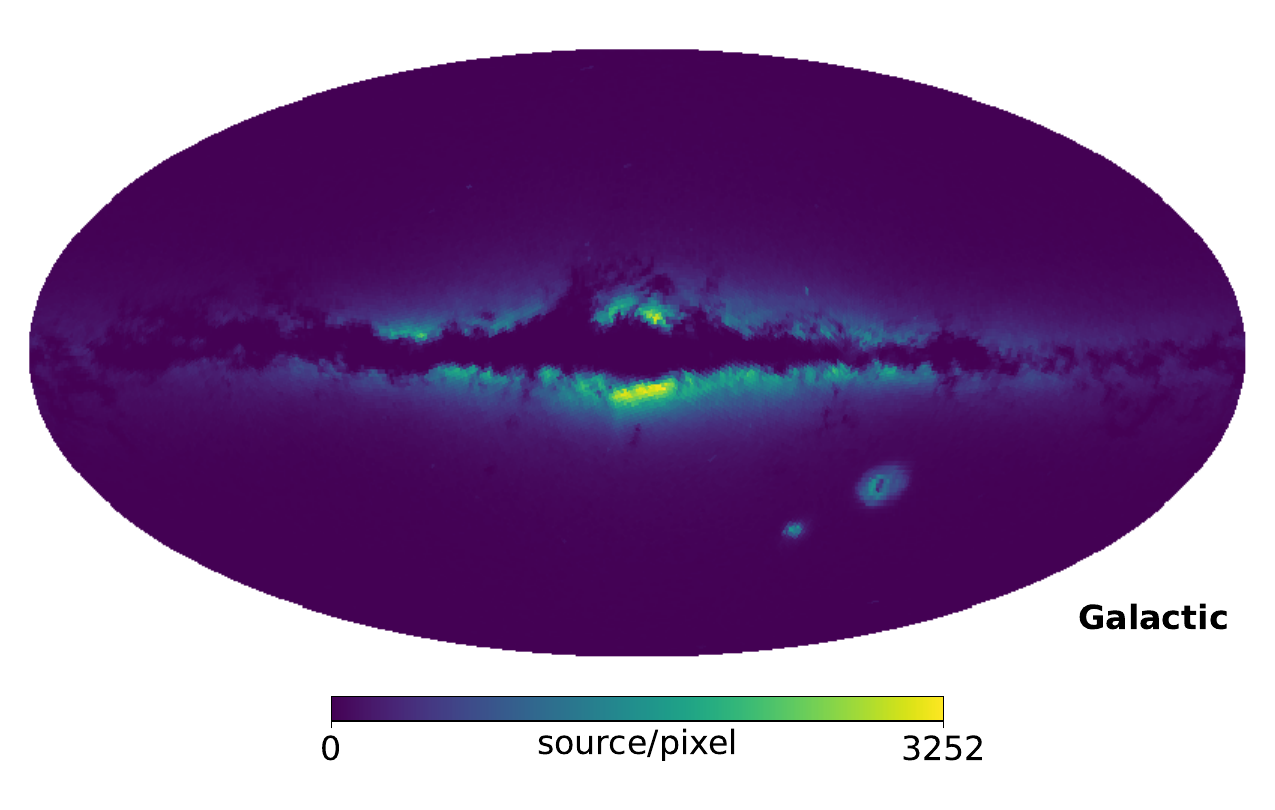}
    \caption{Mollweide map (Galactic coordinates) of the source counts per
HEALPix pixel (\texttt{NSIDE} = 64) for the faint-star sample of
6,465,890 stars, selected from Gaia EDR3 following the criteria of
\citet{2024MNRAS.52710937Y} and used as a stellar contamination template in
Sec.~\ref{sec:cross-correlate}.}
    \label{fig:star_map}
\end{figure}

Nevertheless, the stellar density near the Galactic plane Fig.~\ref{fig:star_map} risks contaminating the quasar sample with misclassified stars, whose non-negligible proper motions could bias the recovered signal. We therefore repeated all analyses with two binary HEALPix masks that exclude pixels whose centres lie at Galactic latitude $|b| \leq 20^\circ$ and $|b| \leq 30^\circ$, removing 34.14\% and 49.98\% of the sky, respectively. Pixels containing no sources are additionally excluded (Sect.~\ref{sec:map_making}). The same masks are applied consistently in the pseudo-$C_\ell$ estimation (Sect.~\ref{sec:searching_systematics}) in the SBI forward model (Sect.~\ref{sec:sbi}), and we investigated the effect of masking on both the dipole and the power spectrum.

\subsection{Quaia} 
We also used the Gaia-unWISE Quasar Catalog, Quaia \citep{Storey_Fisher_2024} as our quasar sample. Quaia is an all-sky and highly homogeneous catalogue, with 1,295,502 sources with the magnitude limit of $G < 20.5$. It is derived from the quasar candidates identified in Gaia DR3 combined with mid-infrared information from unWISE catalogue from Wide-field Infrared Survey Explorer (WISE) observation \citep{ 2010AJ....140.1868W,  Meisner_2019} to improve the sample selection and redshift information. The redshift is estimated via Gaia and unWISE photometry, Gaia-estimated spectral redshift, and cross-matched with Sloan Digital Sky Survey (SDSS) DR16 \citep{2020ApJS..250....8L} with high-precision spectroscopic redshifts. 
However, at fainter magnitudes, the catalogue becomes less reliable due to degraded measurement precision and stronger systematic effects. Therefore, we used a brighter sample with G < 20.0, rather than G < 20.5, with 755,850 sources for a cleaner sample. Since Quaia does not provide inter-component correlations directly, we cross-matched against Gaia DR3 to retrieve the proper motion correlation coefficient $\rho_\mu$ for each source.
We then cross-matched the resulting sample against Gaia EDR3 using \texttt{source\_id}, resulting in 634,401 sources with complete astrometric covariance information, which were used in the map-making procedure described in Sec.~\ref{sec:map_making}. 
\subsection{Map-making}\label{sec:map_making}

We pixelize the individual source measurements from both catalogues into HEALPix maps of resolution \texttt{NSIDE = 64}, corresponding to a pixel size of approximately $0.92^\circ$. This compresses the raw source catalogues into a much smaller set of pixels and speeds up the subsequent computations. For each pixel $p$ containing $N_p$ sources with proper motion ($\mu_{\alpha*,\,i},\,\mu_{\delta,\,i}$), uncertainties ($\sigma_{\mu_{\alpha*,\,i}},\,\sigma_{\mu_{\delta,\,i}}$), and inter-component correlation $\rho_{\mu,\,i}$, the pixel-inverse covariance matrix is first accumulated as the sum of individual per-source inverse covariance matrices: 
\begin{equation}
\Sigma_p^{-1} = \sum_{i \in p} \Sigma_i^{-1}
\label{eq:inverse_pixel_covariance}
\end{equation}
where $\Sigma_i^{-1}$ is the $2\times 2$ inverse covariance matrix of source i
\begin{equation}
{\Sigma}_i^{-1} = 
\dfrac{1}{1-\rho_{\mu,\,i}^2}\begin{bmatrix}
\sigma^{-2}_{\mu_{\alpha^*},\,i} & -\rho_{\mu,\,i} \,(\sigma_{\mu_{\alpha^*},\,i} \,\sigma_{\mu_{\delta},\,i})^{-1} \\
-\rho_{\mu,\,i}\, (\sigma_{\mu_{\alpha^*},\,i} \,\sigma_{\mu_{\delta},\,i})^{-1} & \sigma^{-2}_{\mu_{\delta},\,i}
\end{bmatrix}
\label{eq:inverse_source_covariance}
\end{equation}
The pixel level covariance was obtained by inverting $(\Sigma_p)^{-1}$, and the inverse-variance weighted-mean proper motion of pixel $p$ is
\begin{equation}
    \boldsymbol{\mu}_p = \Sigma_p \sum_{i \in p} \Sigma_i^{-1} \boldsymbol{\mu}_i
\end{equation}
The effective pixel-level uncertainties $\sigma_{\mu_{\alpha *},\,p}$,  $\sigma_{\mu_{\delta},\,p}$, and inter-component correlation $\rho_{\mu,\,p}$ are extracted from the diagonal and off-diagonal elements of $\Sigma_p$ respectively. We assigned zero weight to pixels containing no sources and excludeed them from all subsequent via a binary mask $\nu$. The resulting pixelised proper motion maps and their associated covariance maps are used in both the angular power spectrum analysis in Sec.~\ref{sec:searching_systematics} and the inference in Sec.~\ref{sec:inference}. 

\section{Spherical harmonics} \label{sec:sphe_harmo}

Any sufficient smooth function on the sphere can be expanded in scalar spherical harmonics $Y_{\ell m}(\alpha,\, \delta)$ of degree $\ell$ and order $m$, which form a complete orthonormal basis for spin-0 fields. A spin-0 field is one whose value at each point on the sphere is a scalar invariant under local rotations of the coordinate frame. It admits the following expansion: 
\begin{equation}
    f(\alpha,\,\delta) = \sum_{\ell = 0}^{\ell_{max}} \sum_{m = -\ell}^{\ell} a^0_{\ell m}\, Y_{\ell m}(\alpha,\,\delta)
\end{equation}
where the coefficients $a^0_{\ell m }$ is the angular power at each scale $\ell$. In this work, the stellar density map and $\rm M_{10}$ map are spin-0 scalar fields of this form. 

The proper motion field 
\begin{equation}
    \mathbf{\mu}(\alpha,\,\delta) = \mu_{\alpha*} \,\mathbf{e}_{\alpha} +\mu_{\delta} \,\mathbf{e}_{\delta} 
\end{equation}
however, is not the scalar field. It is a spin-1 vector field, as it assigns a tangent vector at each point on the sphere.

Unlike scalar fields, its expansion separates into a curl-free component (E mode) and a divergence-free component (B mode), resembling the electric field and the magnetic field, respectively. The proper motion field is accordingly decomposed as

\begin{equation}
\boldsymbol{\mu}(\alpha, \delta) = \sum_{\ell =1}^{\ell_{max}} \sum_{m=-\ell}^{\ell} \left(a^E_{\ell m} \boldsymbol{S}_{\ell m} + a^B_{\ell m} \boldsymbol{T}_{\ell m} \right).
\label{eq:mu_function}
\end{equation}
Here $\boldsymbol{S}_{\ell m}$ and $\boldsymbol{T}_{\ell m}$ are the spheroidal (E-mode) and toroidal (B-mode) VSHs, of degree $\ell$ and order $m$ \citep{2012A&A...547A..59M}. The terms $a^E_{\ell m}$ and $a^B_{\ell m}$ are the coefficients describing the magnitude of the spheroidal and toroidal components, respectively. Specifically, the first-order harmonics of spheroidal harmonics $\boldsymbol{S}_{1 m}$ have similar mathematical expression to the effect of acceleration given by Eq.~\ref{eq:false_pm} (Sect 4.2 in \citet{2012A&A...547A..59M}). The acceleration components are then given by 
\begin{equation}
    g_x = -a_{11}^{E,\,R}\sqrt{\dfrac{3}{4\pi}},\quad
    g_y = a_{11}^{E,\,I}\sqrt{\dfrac{3}{4\pi}}, \quad
    g_z = a_{10}^E\sqrt{\dfrac{3}{8\pi}}
\label{eq:convert_dipole}
.\end{equation} 
The superscript R and I here represent the real and imaginary part of the corresponding complex quantities.  In our analysis, we performed the same VSH decomposition on the observed proper motion field.

\section{Searching for systematics with angular power spectra} \label{sec:searching_systematics}
While VSH decomposition in Sec.~\ref{sec:sphe_harmo} extracts the acceleration signal from the proper motion field, it does not characterise the full angular structure of the field to diagnose the presence of the systematics. To address this, we employed the angular power spectrum formalism, which quantifies the statistical power of the field at each angular scale. Specifically, we computed the auto-power spectrum of the quasar proper motion field to characterise its full multipole structure, and the cross-power spectra between the proper motion field and systematic templates to test for spurious correlations
\subsection{Angular power spectra estimation}
In the present analysis, two distinct types of fields appear: the quasars and stellar proper motion field $\mathbf{\mu}(\alpha,\,\delta)$ is a spin-1 vector field, whereas the stellar density map and the $\rm M_{10}$ map are spin-0 fields. We computed the auto-spectra and cross-spectra of these fields using pseudo-$C_\ell$ estimator \citep{Alonso_2019}. 
For the $a$ and $b$ observed on the full sky, the power spectrum can be estimated using: 
\begin{equation}
    \hat{C}^{ab}_\ell = \dfrac{1}{2\ell+1}\sum_{m = -\ell}^{\ell}  a_{\ell m} \,   b^*_{\ell m} 
    \label{eq:cl_amp}
\end{equation}
where $a_{\ell m}$ and $b_{\ell m}$ are respectively the spherical harmonic coefficients of field $a$ and $b$, and $b^*$ denotes the complex conjugate of $b$. In practice, observations cover only a fraction of sky defined by binary masks $\nu^a$ and $\nu^b$, with $\nu = 1$ for observed pixels and $\nu = 0$ otherwise. The masks used in this work are described in \ref{sec:gaia}. Therefore, what we actually observe is $\tilde a(\hat n) = \nu^a(\hat n)$ and $\tilde b(\hat n) = \nu^b(\hat n)$.  

Multiplying by a mask in real space is equivalent to a convolution in harmonic space, which mixes power between multipoles \citep{Alonso_2019}. Applying Eq.~\ref{eq:cl_amp} to the masked field gives the pseudo-$C_\ell$: 
\begin{equation}
    \tilde C^{ab}_{\ell} = \dfrac{1}{2\ell + 1} \sum_{m = -\ell}^{\ell} \tilde a_{\ell m} \, \tilde b^*_{\ell m}
\end{equation}
This quantity is a biased estimate of the true $C^{ab}_\ell$. It ensemble average can be written as \citep{2002ApJ...567....2H, 2005MNRAS.360.1262B}:
\begin{equation}
    \braket{    \tilde C^{ab}_\ell} = \sum_{\ell'} M^{ab}_{\ell \ell'}C^{ab}_{\ell'} 
    \label{eq:ens_avg}
\end{equation}
where $M^{ab}_{\ell \ell'}$ is called mode-coupling matrix that depends on the masks of both fields. The unbiased estimate of $\hat{C}^{ab}_\ell$ is recovered by inverting the coupling matrix:

\begin{equation}
    \hat C_{\ell}^{ab}
    = \sum_{\ell'} (M^{ab})^{-1}_{\ell \ell'}\, \tilde C_{\ell'}^{ab}
    \label{eq:pseudo-cl}
\end{equation}

The variance of the $\hat{C}_\ell$ is estimated as the standard deviation of 500 Monte Carlo realisations of Gaussian random fields, drawn from the measured band-powers $\hat{C}_\ell^{EE}$ and $\hat{C}_\ell^{BB}$ of the proper motion field as the fiducial power spectrum (and $\hat{C}_\ell$) of each scalar template for cross-spectra). For each realisation, synthetic spin-1 proper motion maps are generated via \texttt{healpy.alm2map\_spin} and scalar maps via \texttt{healpy.synfast}, passed through the identical \texttt{NaMaster} pseudo-$C_\ell$ pipeline with the same sky mask, and the resulting decoupled band-powers are recorded. 

In this work, we computed three types of power spectra. (i) the auto-power spectrum of the spin-1 quasars proper motion field, obtained by setting $a=b$ in Eq.~\ref{eq:pseudo-cl}; (ii) the cross-power spectrum between two spin-1 fields, namely the quasar proper motion field and the faint stellar proper motion field; and (iii) the cross-power spectrum between the spin-1 proper motion field and a spin-0 scalar systematic template $T$. For the auto-correlation of spin-1 proper motion field, the pseudo-$C_\ell$ estimator yields four spectra, which are $\hat{C}_\ell^{EE}$, $\hat{C}_\ell^{BB}$, $\hat{C}_\ell^{EB}$, and $\hat{C}_\ell^{BE}$, corresponding to all combinations of E- and B-mode correlations. The same four spectra arise for the cross-correlation between two spin-1 fields. For the spin-1 $\,\times\,$ spin-0 case ~(iii), the estimator yields two spectra, which are $\hat{C}_\ell^{E\times T}$ and $\hat{C}_\ell^{B\times T}$. In the absence of systematic correlations between the two fields, all cross-spectra are expected to be consistent with zero. Any significant detection therefore signals a spurious correlation introduced by instrumental or astrophysical systematics.

\subsection{Auto-power spectra of the proper motion field}

\begin{figure*}[t]
  \sidecaption
  \includegraphics[width=12cm]{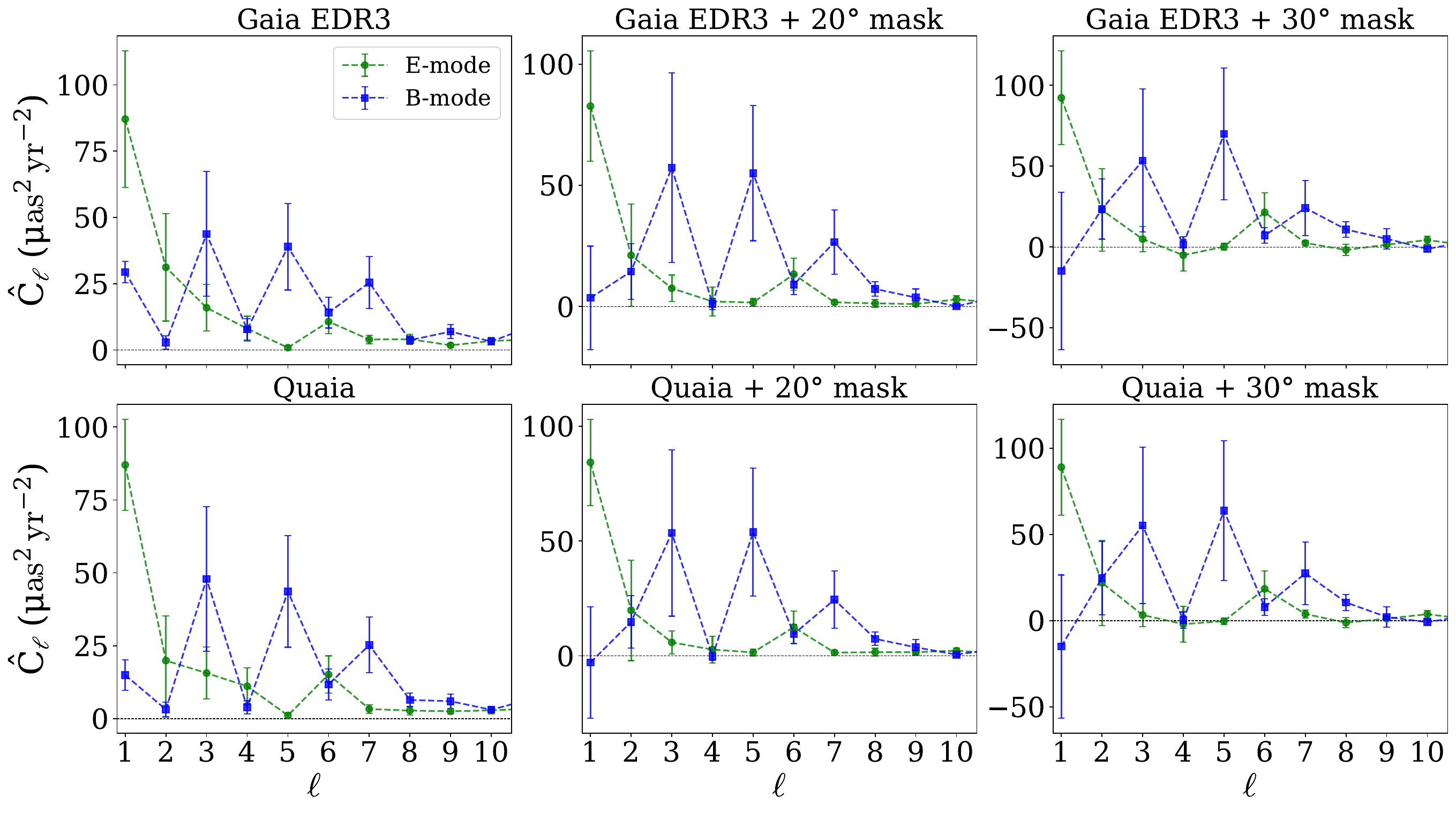}
  \caption{Auto-power spectra ($\hat{C}_{\ell}$) of the E mode and B mode in the proper motions of the quasar sources. The error bars represent the standard deviation across 500 Monte Carlo realisations. The green dashed line represents the power of the E mode of the field, while the blue dashed line represents that of the B mode.}
  \label{fig:cl_auto}
\end{figure*}

The auto-power spectra $\hat{C}_\ell$ for the E mode and B mode of the proper motion field are shown in Fig.~\ref{fig:cl_auto} for all six catalogue configurations, with the corresponding signal-to-noise ratios discussed in Appendix~\ref{app:snr_auto_spectra}. The B-mode power is systematically higher than E-mode power in odd values of $\ell$ and more similar for even values of $\ell$. This even-odd E-B asymmetry has been attributed to the Gaia scanning strategy \citep{2022A&A...667A.148G}, a hypothesis we examine through cross-correlation in Sec.~\ref{sec:cross-correlate}.

A notable feature of Figs.~\ref{fig:cl_auto} and \ref{fig:snr_auto_per_ell} is a large dipole B mode with the signal-to-noise ratio of $\sim 7$ for unmasked Gaia EDR3. Since the secular aberration drift produces only an E-mode dipole and cannot generate B-mode power. A B-mode dipole instead represents a residual spin of the quasar reference frame, which is a documented, magnitude-dependent property of the Gaia astrometric solution \citep{2021A&A...649A.124C,2021A&A...649A...2L}. However, this spin does not bias the acceleration, as the E- and B-mode dipoles are orthogonal on the full sky, so the frame rotation does not project onto $\mathbf{g}$. With incomplete sky coverage these modes can couple, but the toroidal coefficients $a^{B}_{\ell m}$ are free parameters, and every forward simulation applies the same coverage, weighting, and mask as the data. Therefore, this coupling is reproduced in the simulations and marginalised over in the fit (Sec.~\ref{sec:inference}), leaving the recovered acceleration robust.

We also note that some slightly negative band-powers appear at certain multipoles in the $20^\circ$ and $30^\circ$ mask configurations. While the true auto-power spectrum of a field is positive by definition, the pseudo-$C\ell$ estimate $\hat{C}_\ell$ need not be so\footnote{\url{https://namaster.readthedocs.io/en/latest/1BasicFunctionality.html}}.

\subsection{Cross-correlation with systematics templates} \label{sec:cross-correlate}

Quasar proper motions should ideally show no correlation with unrelated fields like a satellite's scanning strategy, assuming the data are devoid of systematic biases. The detection of significant cross-correlations can therefore signal potential errors or biases in the observational strategies or data processing techniques employed. We examined three systematic templates: the $\rm M_{10}$ scanning strategy map, the faint stellar density map, and the faint stellar proper motion map.

\begin{figure}[htbp]
    \centering
    \includegraphics[width=\linewidth]{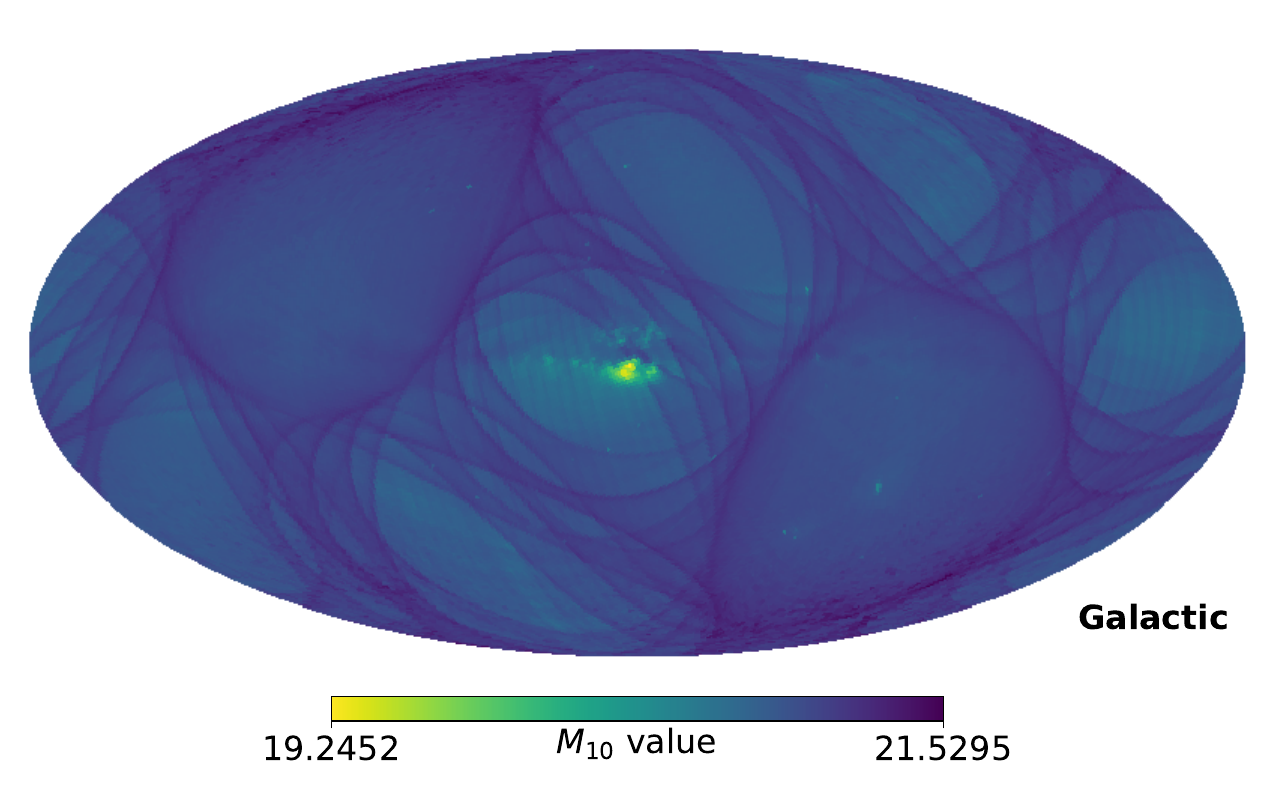}
    \caption{Median magnitude of sources with $\leq 10$ Gaia transits ($\rm M_{10}$)}
    \label{fig:scanning}
\end{figure}

The $\rm M_{10}$ map (is shown in Fig.~\ref{fig:scanning}). This quantity is the median magnitude of source with  $\leq 10$ Gaia transits, which incorporates both scanning strategy and source crowding \citep{2023A&A...669A..55C}. We down-sampled the map to \texttt{NSIDE} = 64.

\begin{figure*}[t!]
  \sidecaption
  \includegraphics[width= 12cm]{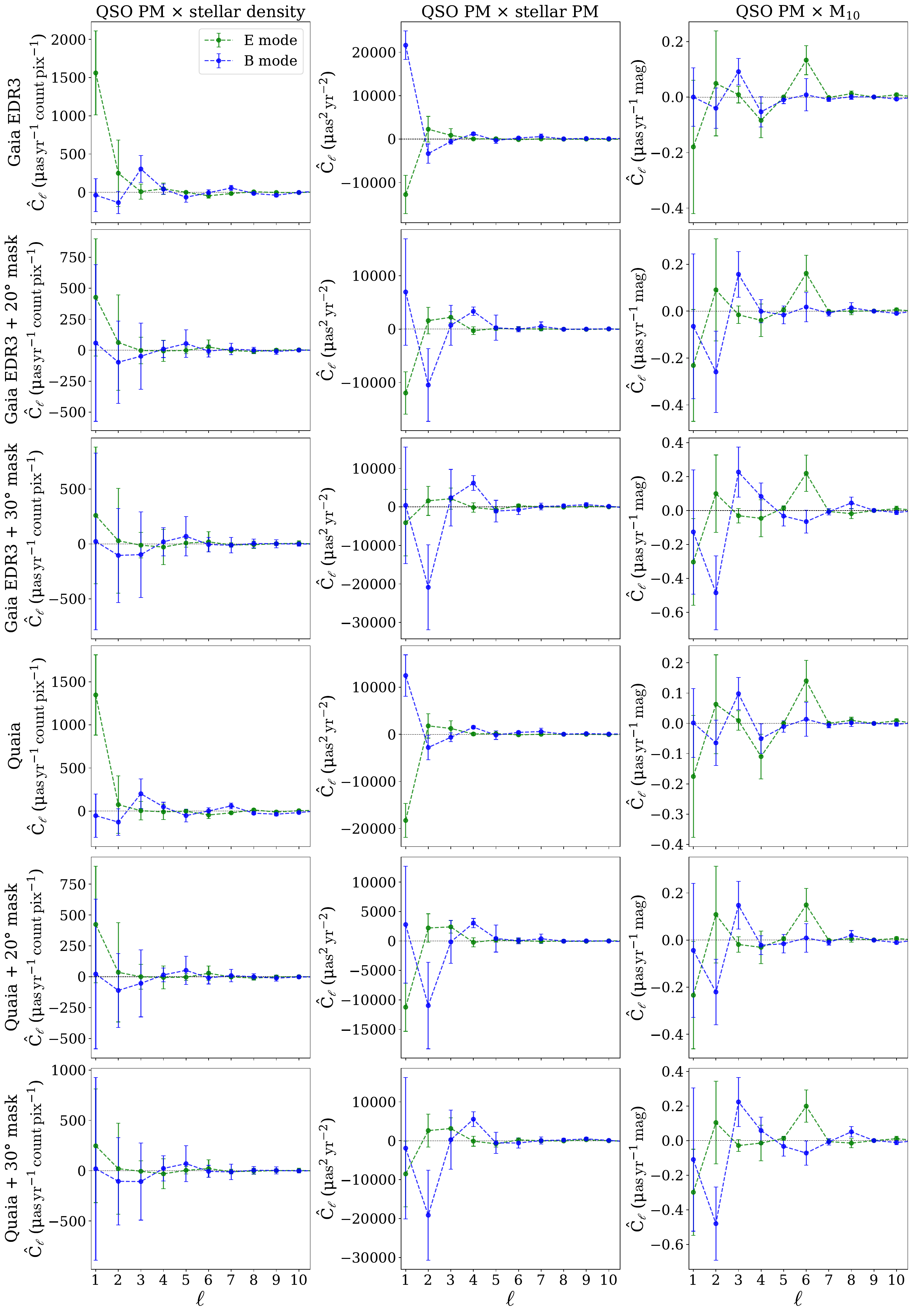}
  \caption{Cross-power spectra $\hat{C}_{\ell}$ of the E mode ( dashed green line) and B mode (dashed blue line) between the proper motion field and systematic templates. Each column shows a different template: the first column is the cross-correlation with the faint stellar density map, the second column is the cross-correlation with the faint stellar proper motion map, and the third column is the cross-correlation with the $\rm M_{10}$ map. Each row corresponds to a different catalogue and masking configuration.}
  \label{fig:cross_spectra_panel}
\end{figure*}

The cross-correlation between the proper motion field and the $\rm M_{10}$ map is shown in the third column in Fig.~\ref{fig:cross_spectra_panel}. Both the E mode and B mode cross-power values fluctuate around zero with no persistent signal at any multipole. The signal-to-noise ratios remain within $2\sigma$ across all multipoles for every sample and mask configuration (see Fig.~\ref{fig:snr_per_ell} and Appendix~\ref{app:snr_cross_spectra}). The scanning strategy therefore shows no statistically significant correlation with the quasar proper motion field, in contrast with the suggestion of \citet{2022A&A...667A.148G} that the scanning law drives the even-odd E-B asymmetry.

The catalogue of faint stars was constructed from EDR3 with the constraints outlined in \cite{2024MNRAS.52710937Y}, resulting in 6,465,890 stars. These faint stars are the most likely to be confused with quasars. The cross spectrums are shown in the first column in Fig.~\ref{fig:cross_spectra_panel}. One can see that for the unmasked Gaia EDR3 sample, a strong E-mode signal $\ell = 1$, of order $\sim 1500 \,\mu\mathrm{as} \,\mathrm{yr}^{-1} \,\mathrm{count}\, \mathrm{pix}^{-1}$, indicating a significant dipole-scale correlation between proper motion field and spatial distribution of faint stars. A similar signal is present for Quaia with power of order $\sim 1300 \,\mu\mathrm{as} \,\mathrm{yr}^{-1} \,\mathrm{count}\, \mathrm{pix}^{-1}$ at $\ell =1$. For the unmasked Gaia EDR3 and Quaia samples, a strong E-mode signal at $\ell = 1$ is present, indicating a significant dipole-scale correlation between the proper motion field and the spatial distribution of faint stars. However, this signal is suppressed when applying the galactic masks (see Appendix~\ref{app:snr_cross_spectra}), which shows that this correlation is driven by Galactic plane contamination. There is no significant correlation at other multipoles across configurations since all signals are within $2 \sigma$. 

As an additional systematic diagnostic, we cross-correlated the quasar proper motion field with the proper motion field of the faint stellar sample. The cross-spectra are shown in the second column in Fig.~\ref{fig:cross_spectra_panel}. A large amplitude signal is present at $\ell =1$ for the unmasked Gaia EDR3 and Quaia samples in both E mode and B mode, with opposite sign between the E mode and B mode. We, again, attribute this to Galactic plane contamination. Applying the galactic masks substantially suppresses this signal across all multipoles.

\section{Simulation-based inference} \label{sec:sbi}

Based on Sec.~\ref{sec:aberration} and Sec.~\ref{sec:sphe_harmo}, we know that we can infer the acceleration of the Solar System from the VSH decomposition of the proper motion field. From Sec.~\ref{sec:searching_systematics}, we can see that the proper motion data contain unknown large-scale noise; therefore, it is reasonable to include these features into the model for a robust inference. The first step is to extract the VSH coefficients from the proper motion field, which is described in Sec.\ref{sec:solver}. The model and the inference process are described in Sec.~\ref{sec:inference}

\subsection{Solver} \label{sec:solver}
To extract VSH coefficients from the observed proper motion field, we solved a weighted least-squares problem that accounts for the full per-pixel covariance between two proper motion components. For each unmasked pixel $p$, the observation is a two-component vector $\mathbf{d}_p = (\mu_{\alpha^*,p}, \,\mu_{\delta,p})^T$ with the 2x2 inverse covariance matrix $(\Sigma_p)^{-1}$ in Eq.~\ref{eq:inverse_pixel_covariance}.
The forward model maps a set of VSH coefficients ($\mathbf{x} = {a_{\ell m}^E, a_{\ell m}^B}$) (for $1\leq \ell \leq \ell_{max}$) to a predicted proper motion field on the sphere via a linear operator \textbf{A}:
\begin{equation}
\mathbf{d}^{\mathrm{model}} = \mathbf{A}\,\mathbf{x}
\end{equation}
Applying \textbf{A} is equivalent to an inverse spin-1 transform evaluated on the pixel grid\footnote{Implemented with \texttt{healpy}'s \texttt{alm2map\_spin} routine \citep{Zonca2019}; the adjoint transform in the normal equations is the corresponding adjoint synthesis operation.} that relates the E-mode and B-mode coefficients to the proper motion components through $\mu_{\alpha^*} = m_\phi$, and $\mu_{\delta} =- m_\theta$, where $m_\phi$ and $m_\theta$ are the standard spin-1 harmonic field components computed by \texttt{healpy}.

The optimal estimate of \textbf{x} is obtained by solving  
\begin{equation}
    \mathbf{M}\,\mathbf{x} = \mathbf{b}
\end{equation} \label{eq:vsh_fit}
where 

\begin{equation}
   \mathbf{M} = \mathbf{A}^T\,\mathbf{C}^{-1} \mathbf{A},\quad \mathbf{b} = \mathbf{A}^T \,\mathbf{C}^{-1}\,\mathbf{d}
\end{equation}

and $\mathbf{C}^{-1} = \mathrm{diag}(\Sigma_{p,\,1}^{-1},\,\Sigma_{p,\,2}^{-1},\,...)$ is the block-diagonal inverse covariance over all unmasked pixels. We never formed $\mathbf{A}$ explicitly. Instead, we implemented the action of the operator $\mathbf{x} \mapsto \mathbf{A}^T\,\mathbf{C}^{-1}\,\mathbf{A}\,\mathbf{x}$ as a three-step sequence: (i) a forward spin-1 spherical harmonic transform to obtain proper motion maps, (ii) pixel-wise multiplication by the full $2\times2$ inverse covariance ${\Sigma}_p^{-1}$, and (iii) adjoint spin-1 spherical harmonic transform to project back to harmonic space. Because this sequence is linear, it defines the matrix ($\mathbf{M}$) implicitly. We built $\mathbf{M}$ one column at a time by setting the $i$-th coefficient to one and all others to zero, then running the three steps above. This returns the $i$-th column of $\mathbf{M}$. Repeating this for every coefficient gives the full matrix. The right-hand side \textbf{b} is computed by applying steps (ii) and (iii) to the observed data instead of model maps. The matrix (\textbf{M}) is a square matrix of $2(\ell_{max} +1)^2 - 2$, which is the number of parameters of E and B with real value for $\ell \geq 1$. The system $\mathbf{M\,x} =\mathbf{b}$ is then solved via LU factorisation, with the factorisation cached and reused across all forward simulation calls to reduce computational cost.

\subsection{Simulation-based inference} \label{sec:inference}

\begin{figure*}[t]
  \sidecaption
  \includegraphics[width= 12 cm]{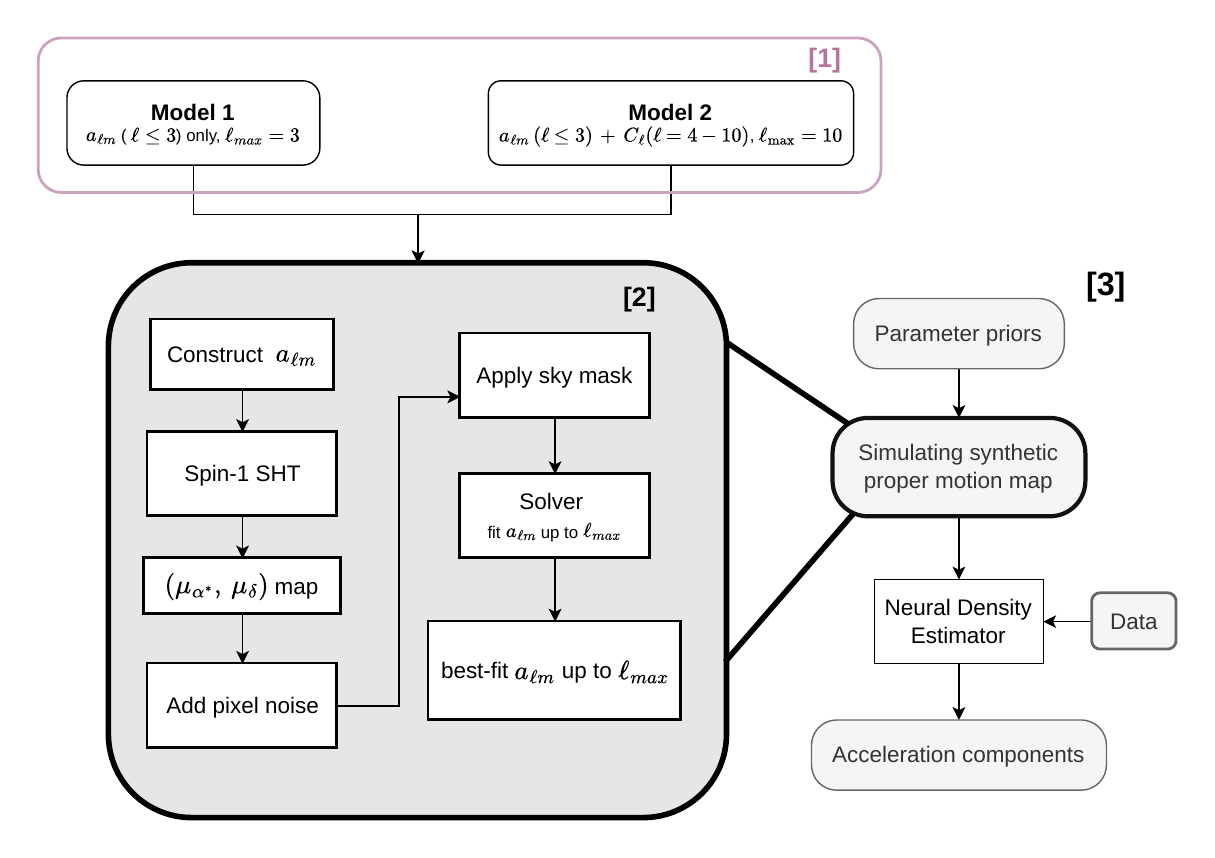}
  \caption{Schematic of the SBI pipeline. It is organised in three sections: the first [1] defines the models (Model 1 and Model 2). The second [2] shows the forward simulation loop (repeated 100{,}000 times). The third [3] shows the overall SBI structure.}
  \label{fig:flowchart}
\end{figure*}
The goal of this work is to infer the acceleration of the Solar System from the observed quasar proper motion field. We frame this as a Bayesian inference problem. 

In Bayesian inference, the goal is to calculate the posterior probability for the observed data $x$ and a given prior, $p(\theta)$, of the input parameters, $\theta$: 
\begin{equation}
    p(\theta\,|\,x) = \dfrac{p(x\,|\,\theta) \; p(\theta)}{\int p(\,x|\, \theta ') \; p(\theta ') \; d\theta '}
    \label{eq:bayes}
\end{equation}

In this work, writing down the likelihood $p(x\,|\,\theta)$ analytically is difficult since we did not know exactly the systematics. Rather than attempt to derive and evaluate this likelihood, we adopted a SBI framework (using \texttt{sbi}\footnote{\url{https://sbi.readthedocs.io/en/latest/index.html}} package; \citet{Tejero-Cantero2020}), or so-called likelihood-free inference, to approximate the posterior, which requires only the ability to simulate mock data from a forward model. The SBI in our pipeline is using the neural density estimator (NDE), which is based on a normalising flow \citep{tabak2010density, https://doi.org/10.1002/cpa.21423, 2015arXiv150505770J} that learns a flexible, invertible mapping between a complex target distribution and a simple probability distribution, typically a multivariate Gaussian. In this case, the neural network defines a flexible and bijective transformation: $f: \mathbf{x} \rightarrow \mathbf{z}$, where $\mathbf{x}$ represents the samples of the target distribution and $\mathbf{z}$ follows the simple base distribution $\pi(\mathbf{z})$. The transformation $f$ is defined to be invertible and its Jacobian determinant can be computed efficiently, the probability density of the original data can be obtained exactly using the change-of-variables formula. Consequently, although the target distribution may be highly complex, its density can be evaluated efficiently by transforming the data into the latent space, evaluating the simple base distribution, and accounting for the volume change introduced by the transformation. The framework is shown in Fig.~\ref{fig:flowchart}.

We defined two forward models to create mock quasar proper motion maps in terms of parameters:

Model~1: $\theta_1 = \{a_{\ell m} (\ell \leq 3)\}$. The model draws only the deterministic VSH coefficients for $\ell = 1-3$, and the solver fits up to $\ell _{max} =3$. The simulated maps contain only low-multipole signal plus pixel noise. This model mirrors the analysis of \citet{2021A&A...649A...9G} where the proper motion field is assumed to consist of dipole plus uncorrelated noise. Since they take the result at $\ell_{max} =3$, we decided to take the deterministic part until that for a robust comparison.  

Model~2: $\theta_2 = \{a_{\ell m}$ ($\ell \leq 3$) + $\hat{C}_\ell$ ($\ell = 4$--$10)\}$. This is an extended version of Model 1 by adding the angular power spectrum $\hat{C}_\ell$ at higher multipoles ($\ell \geq 4$).  Physically, the $\hat{C}_{\ell}$ of higher multipoles capture un-modelled correlated noise on large scales. Since for an arbitrary isotropic field
\begin{equation}
    \left< a_{\ell m} \, a_{\ell' m'} \right> = \hat{C}_{\ell} \,\delta_{\ell \ell'} \, \delta_{m m'}
    \label{eq:cl and alm}
\end{equation}
let $\ell = \ell'$ and $m = m'$. Then Eq.~\ref{eq:cl and alm} becomes 
\begin{equation}
    \mathrm{Var}(a_{\ell m}) = \left< |a_{\ell m}|^2\right> = \hat{C}_{\ell} 
\end{equation}
Therefore, we could treat the high multipoles as a Gaussian noise contribution with the covariance matrices coming from their $C_{\ell}$. We assigned uniform priors from -50 to 50 $[\mu\mathrm{as} \,\mathrm{yr}^{-1}]$ for $a_{\ell m}$ and from 0 to 130 $[\mu\mathrm{as}^2 \,\mathrm{yr}^{-2}]$ for $\hat{C}_{\ell}$, an upper limit chosen to cover the measured $C_\ell$ in Fig.~\ref{fig:cl_auto}.

Given a draw of $\boldsymbol{\theta}$, we simulated a full-sky map of proper motion map as follow: (i) For $\ell = 1-3$, we used the drawn $a_{\ell m}$ directly. For Model 1, no power was added beyond $\ell = 3$. For Model 2, we additionally drew random Gaussian realisations at each $(\ell,m)$ with variance $\hat{C}_\ell$. (ii) It was then converted to the proper motion map using \texttt{healpy}, and Gaussian pixel noise was added using a per-pixel covariance matrix (inversion of Eq.~\ref{eq:inverse_pixel_covariance}).
This is a valid way to create pixel noise as discussed in Appendix~\ref{app:pixel_noise}.
(iii) We applied the same sky mask as the data (including the galactic plane cut). The result is a synthetic quasar proper motion map that produces the same proper motion distribution for the two components, as discussed in Appendix~\ref{app:pm_distribution}. 

Each simulated map is then analysed identically to the real data: we applied the weighted VSH solver (Sect. \ref{sec:solver}) to fit the $a_{\ell m}$ with $\ell_{max} = 3$ for Model 1 and $\ell_{max} =10$ for Model 2. The resulting best-fit $a_{\ell m}$ are the summary statistics -- 30 real-valued parameters for Model 1 and 240 for Model 2. By repeating this forward process for $\sim100,000$ draws of $\boldsymbol{\theta}$, we built a training dataset of (parameters, summary) pairs. An NDE was then trained to learn the conditional density $p(\boldsymbol{\theta}|\text{summary})$. For Model 2, the NDE learns to marginalise over the nuisance power encoded in the $\ell \geq 4$ when constraining the multipoles. Finally, we passed the observed data summary into the NDE to obtain the posterior on $\boldsymbol{\theta}$. The acceleration components $g_x$, $g_y$, and $g_z$ were then extracted from the posterior samples of the dipole components of $a_{\ell m}$ using Eq.~\ref{eq:convert_dipole}.

\subsection{Probability coverage}

We assessed the calibration of the posterior generator using a marginal coverage test. For each parameter $\theta_k$, the marginal expected coverage probability (ECP) is defined as
\\
\begin{equation}
    p_\gamma = \mathbb{P}(\theta_{true} \in \,\mathcal{R}_\gamma{(x)})
,\end{equation}
where $\mathcal{R}_\gamma(x)$ is the estimated posterior region containing a $\gamma$-credible set, defined as interval between the $(1-\gamma)/2$ and $(1+\gamma)/2$ quantiles of the marginal posterior for each parameter. $\gamma  \in (0, 1)$ is the nominal credibility level, which we scanned over a uniform grid. We consider the posterior to be well calibrated with respect to ECP when $p_{\gamma} = \gamma$. If $p_\gamma < \gamma$, which is called a sub-identity, it corresponds to over-confidence. Conversely, $p_\gamma > \gamma$ indicates under-confidence, where the credible intervals are unnecessarily wide. We evaluated the ECP using N = 500 test simulations, drawing 500 posterior samples per test point, and computed the marginal probability curves for the acceleration components and all Model 2 parameters. However, normalising flows employed in NDE can exhibit inaccurate density estimation near hard prior boundaries, which artificially inflates overconfidence in the coverage test \citep{2026A&A...710A.385P}. To mitigate this, we restricted the evaluation to the inner 90\% of each parameter's prior range and discarded posterior samples falling outside this region.

\begin{figure}[htbp]
    \centering
    \includegraphics[width=\linewidth]{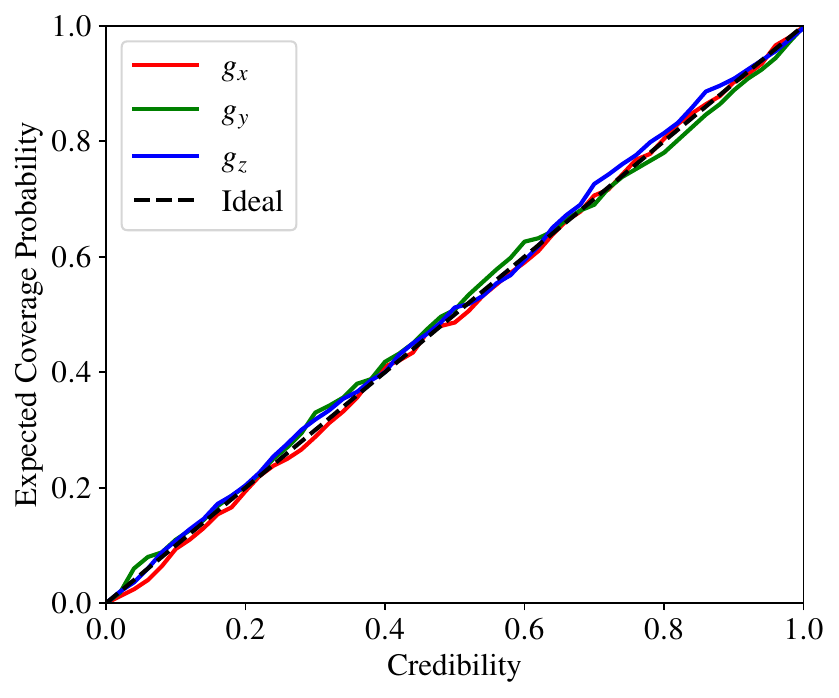}
    \caption{Marginal coverage test for the three acceleration components. A well-calibrated posterior lies on the identity diagonal (dashed black line).}
    \label{fig:tarp_dipole}
\end{figure}

Fig.~\ref{fig:tarp_dipole} shows that the coverage curves of the acceleration components are in good agreement with the identity. This alignment confirms that the NDE produces well-calibrated posteriors for the physically meaningful outputs of Model 2.

\begin{figure}[htbp]

    \centering
    \includegraphics[width=\linewidth]{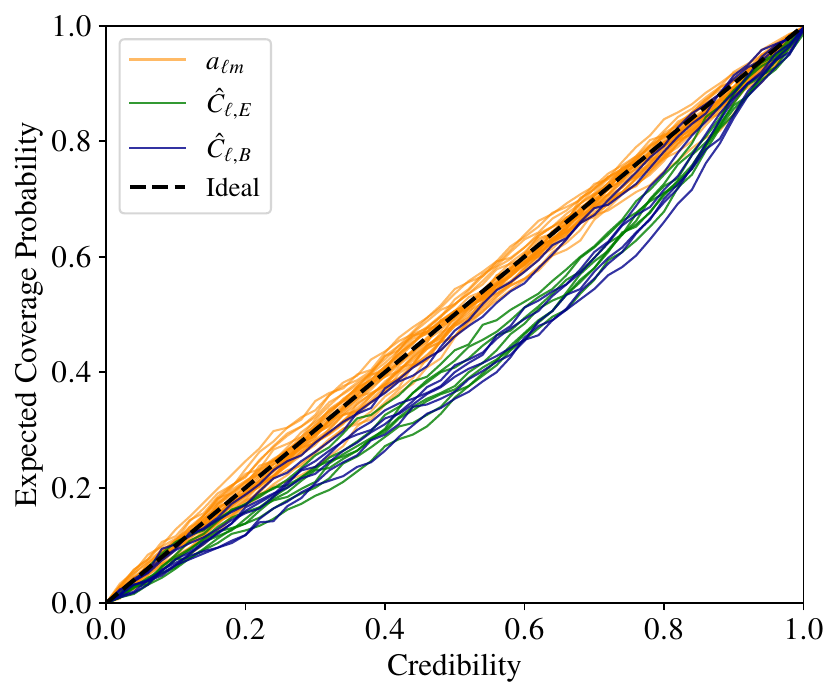}
    \caption{Marginal coverage test for all Model 2 parameters. The ECP $p_\gamma$ is shown for the VSH coefficients (orange line), the E-mode power spectrum amplitude ($\hat{C}_{\ell,\,E}$; green lines), and the B-mode power spectrum amplitude ($\hat{C}_{\ell,\,B}$; (blue lines).}
    \label{fig:tarp_full}
\end{figure}

Figure~\ref{fig:tarp_full} shows the coverage curves for all Model 2 parameters grouped by type ($a_{\ell m},\,\hat{C}_{\ell,\,E}$, and $\,\hat{C}_{\ell,\,B}$). One can see that the $a_{\ell m}$ coefficients are reasonably well calibrated. However, $\hat{C}_\ell$ shows significant deviation from the diagonal.  This can be attributed to the fact that pixel noise is larger than the correlated noise on large-scale modelled by $\hat{C}_\ell$, which shows that the data are largely uninformative about the individual $\hat{C}_\ell$ values. This can be shown in Fig.~\ref{fig:cornet_full_cl}, the posteriors on $\hat{C}_{\ell,\,E}$ and $\hat{C}_{\ell,\,B}$ for $\ell = 4-10$ are broad and approximately uniform across prior range. However, this is expected because the role of $\hat{C}_\ell$ is to marginalise over un-modelled large-scale stochastic power rather than to be individually constrained. The poor individual calibration of these nuisance parameters does not invalidate the acceleration posteriors, which remain well calibrated as demonstrated above.

\subsection{Analysis}

The inference of acceleration of the Solar System is shown in Fig.~\ref{fig:gxyz_corner} and Tab.~\ref{tab:gxyz}. For each sample and masking configuration, we also quantified the consistency of the recovered direction with the Galactic centre using $\sigma_{\rm GC}$, the Gaussian level of a two-degree-of-freedom $\chi^2$ computed from the posterior under the null hypothesis that the acceleration points at the Galactic centre.  Model 1 fits yield consistent acceleration components across all samples and masking configurations. For unmasked Gaia EDR3 sample, we obtain ($g_x,\,g_y,\,g_z$) = $(-0.14^{+0.36}_{-0.36}, \,-4.06^{+0.33}_{-0.31},\,-2.78_{-0.34}^{+0.34}) \;\rm \mu as\, yr^{-1}$, which is in good agreement with \citet{2021A&A...649A...9G}. The Quaia values are similarly consistent with ($g_x,\,g_y,\,g_z$) = $(0.04^{+0.42}_{-0.42}, \,-4.30^{+0.37}_{-0.37},\,-2.69^{+0.38}_{-0.39})\;\rm \mu as\, yr^{-1}$. The acceleration magnitude under Model 1 is $|\mathbf{g}| =$ $4.94^{+0.33}_{-0.31}$$\;\mu\text{as}\,\text{yr}^{-1}$ for Gaia EDR3 and $5.10^{+0.38}_{-0.37}$$\rm \; \mu as \,yr^{-1}$ for Quaia, detected at $15.4\sigma$ and $13.7\sigma$ respectively. The corresponding directions, $(\alpha,\, \delta) =
  \left(267.94^{+5.1}_{-5.1}, \,-34.36^{+4.0}_{-3.8}\right)^\circ$ for Gaia
  EDR3 and $\left(270.55^{+5.6}_{-5.6},\, -32.10^{+4.5}_{-4.1}\right)^\circ$
  for Quaia, lie within $\sim 3\degr$ of the apex found by
  \citet{2021A&A...649A...9G}. The recovered directions show no significant offset with the Galactic centre, with $\sigma_{\rm GC} = 0.89$ and $0.55$, respectively. The detection significance decreases under galactic masking, to $12.0 \sigma$ ($11.0\sigma$) for the $20^\circ$ cut and $7.9\sigma$ ($7.6\sigma$) for the $30^\circ$ cut in Gaia EDR3 (Quaia), while the recovered
  directions remain stable, with directional uncertainties growing to
  $\sim 8$--$14\degr$ (Tab.~\ref{tab:gxyz}).

\begin{figure}

    \centering
    \includegraphics[width=\linewidth]{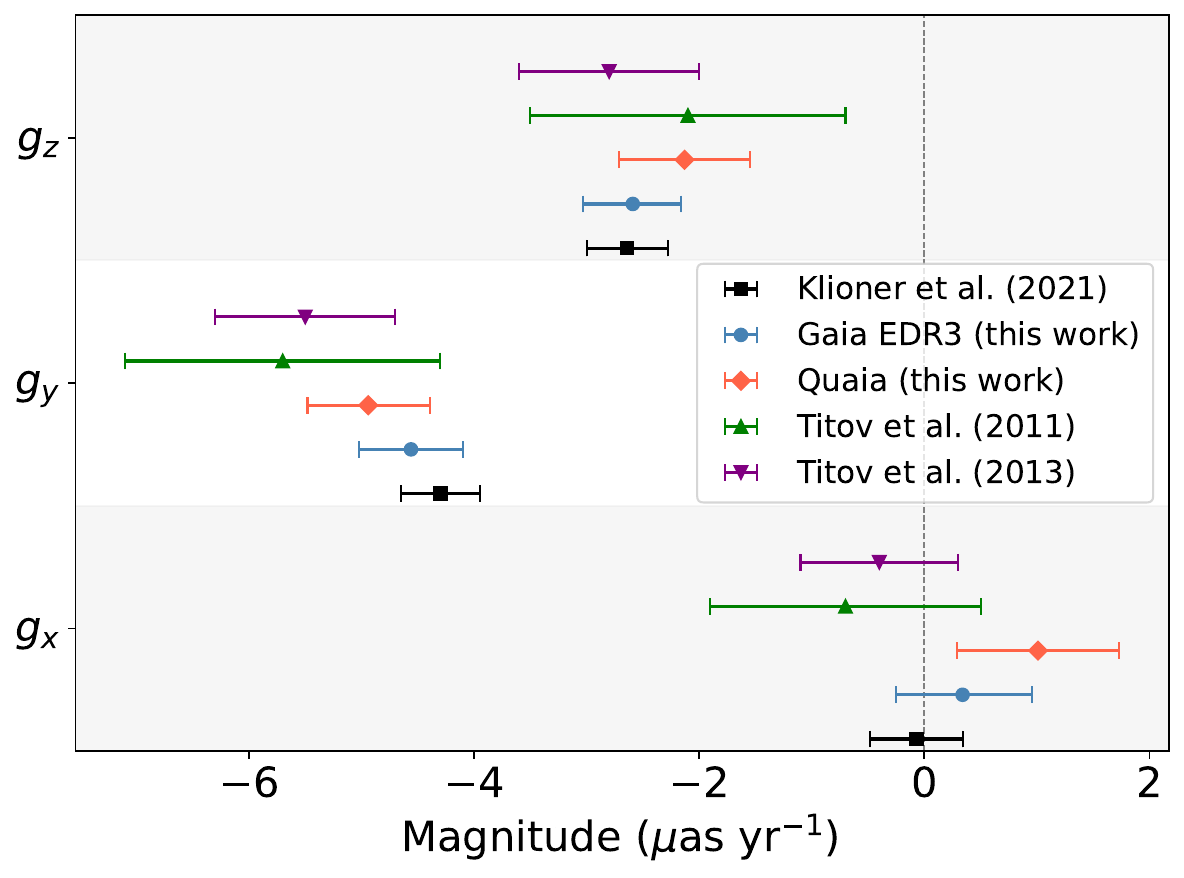}
    \caption{Comparison of acceleration components inferred in this work using Model 2 for the \textit{Gaia} EDR3 quasar catalogue and the Quaia catalogue, against the previous determinations of \citet{2011A&A...529A..91T}, \citet{2013A&A...559A..95T}, and \citet{2021A&A...649A...9G}. Error bars represent $1\sigma$ uncertainties.}
    \label{fig:accel_compare_component}
\end{figure}

\begin{figure*}[htb!]
    \centering

    \subfloat[Gaia EDR3]{
        \includegraphics[width=0.32\linewidth]{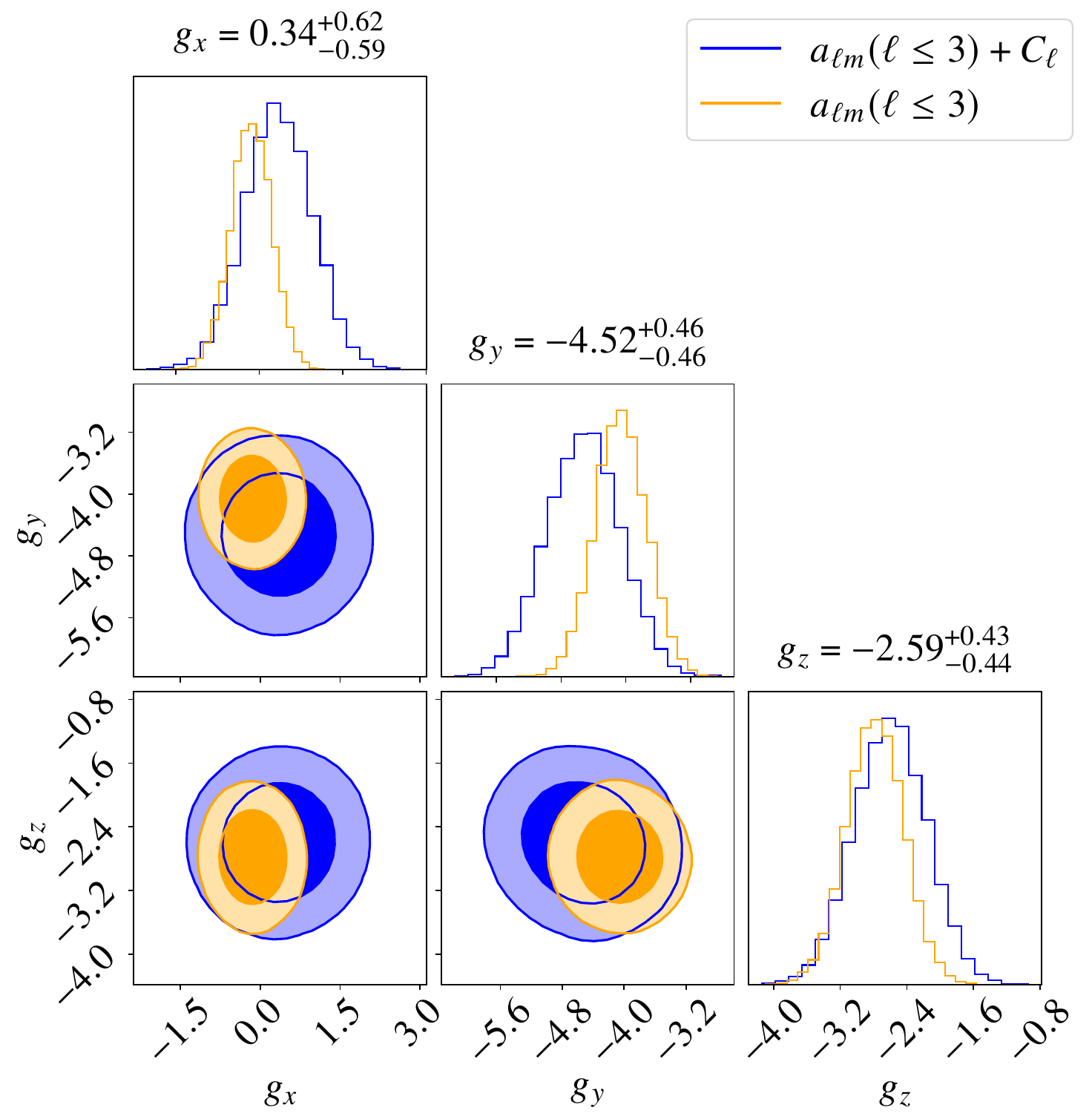}
        \label{subfig:a}
    }\hfill
    \subfloat[Gaia EDR3 + $20^\circ$ galactic mask]{
        \includegraphics[width=0.32\linewidth]{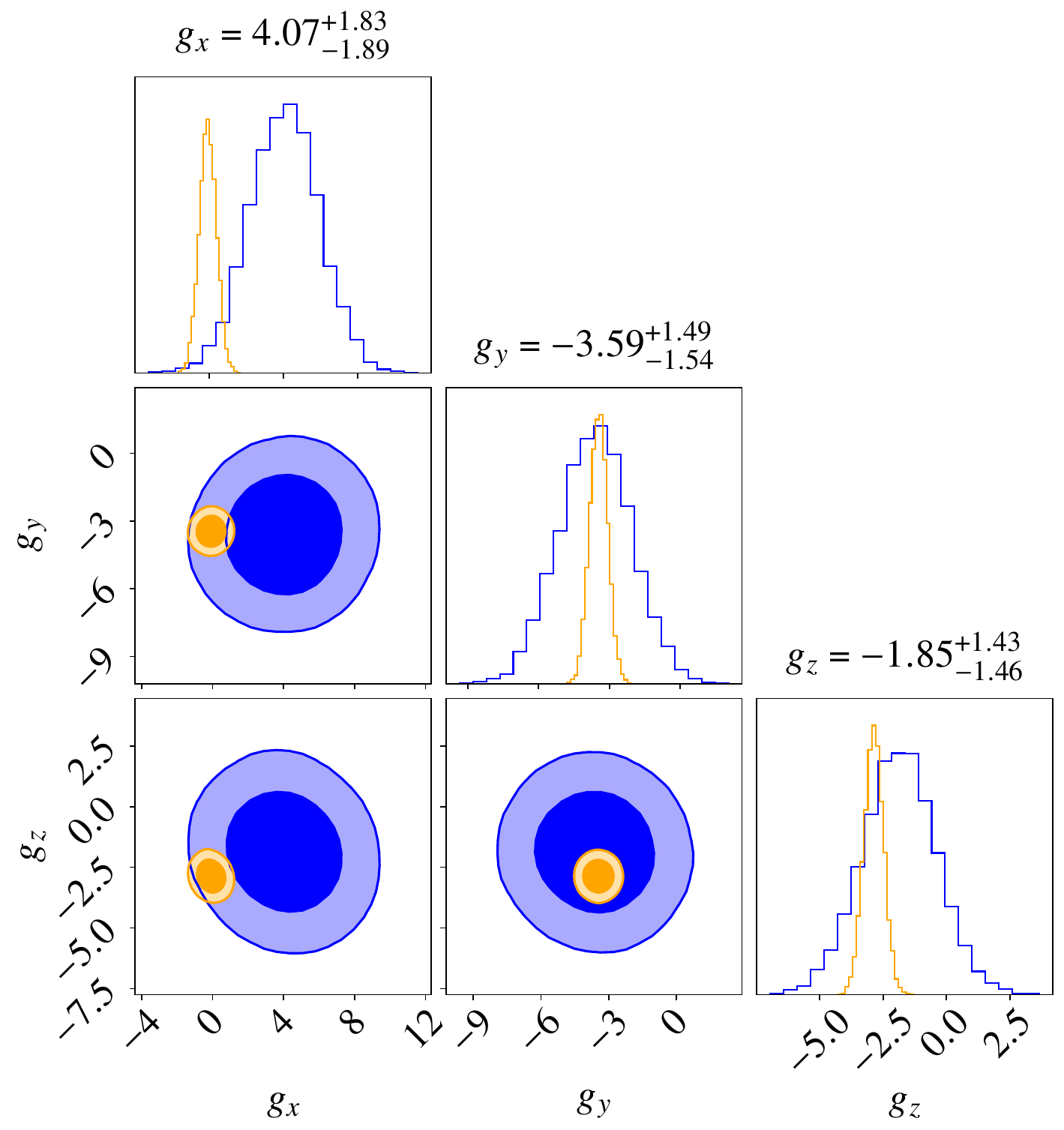}
        \label{subfig:b}
    }\hfill
    \subfloat[Gaia EDR3 + $30^\circ$ galactic mask]{
        \includegraphics[width=0.32\linewidth]{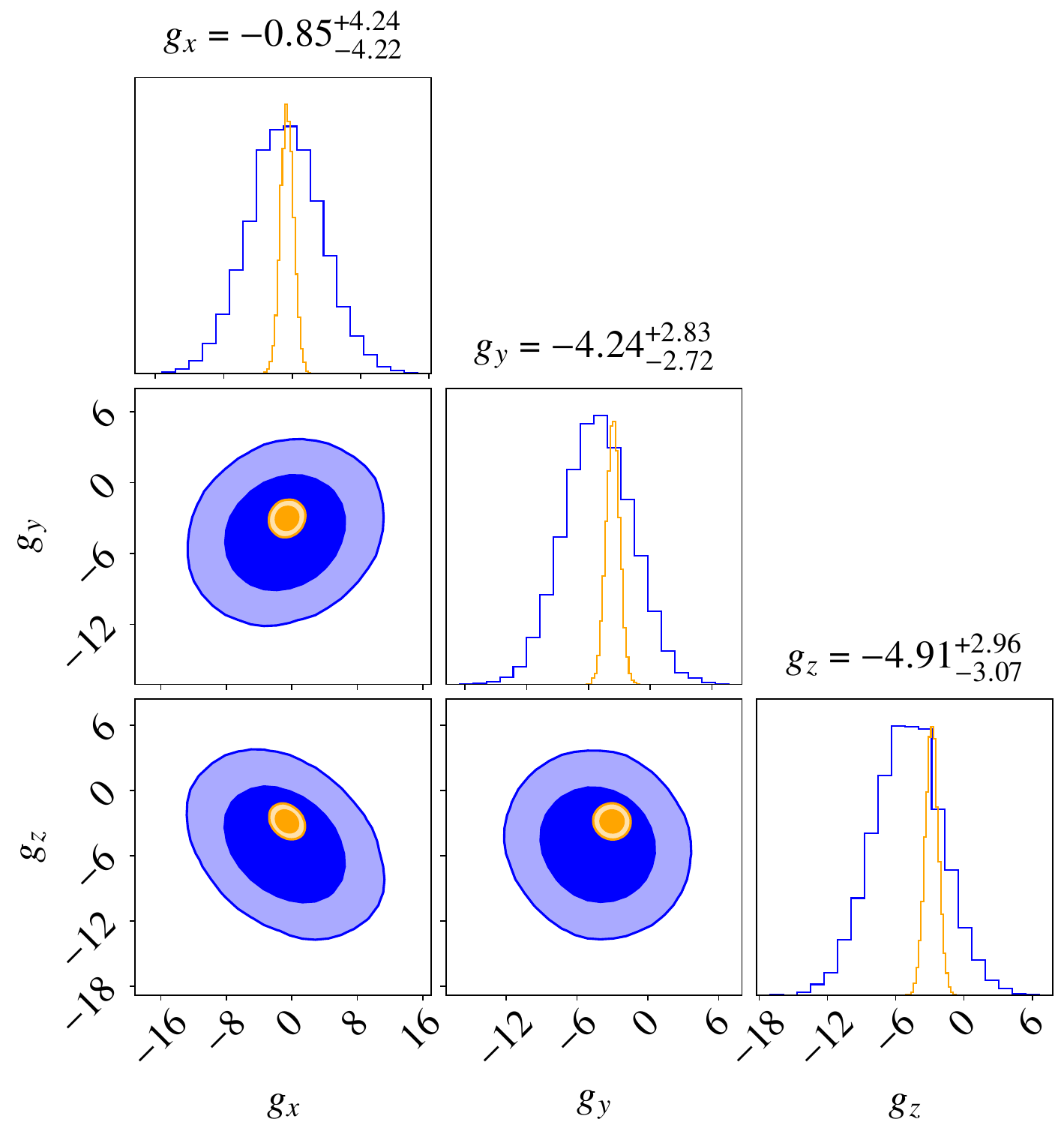}
        \label{subfig:c}
    }\\[0.5em]

    \subfloat[Quaia]{
        \includegraphics[width=0.32\linewidth]{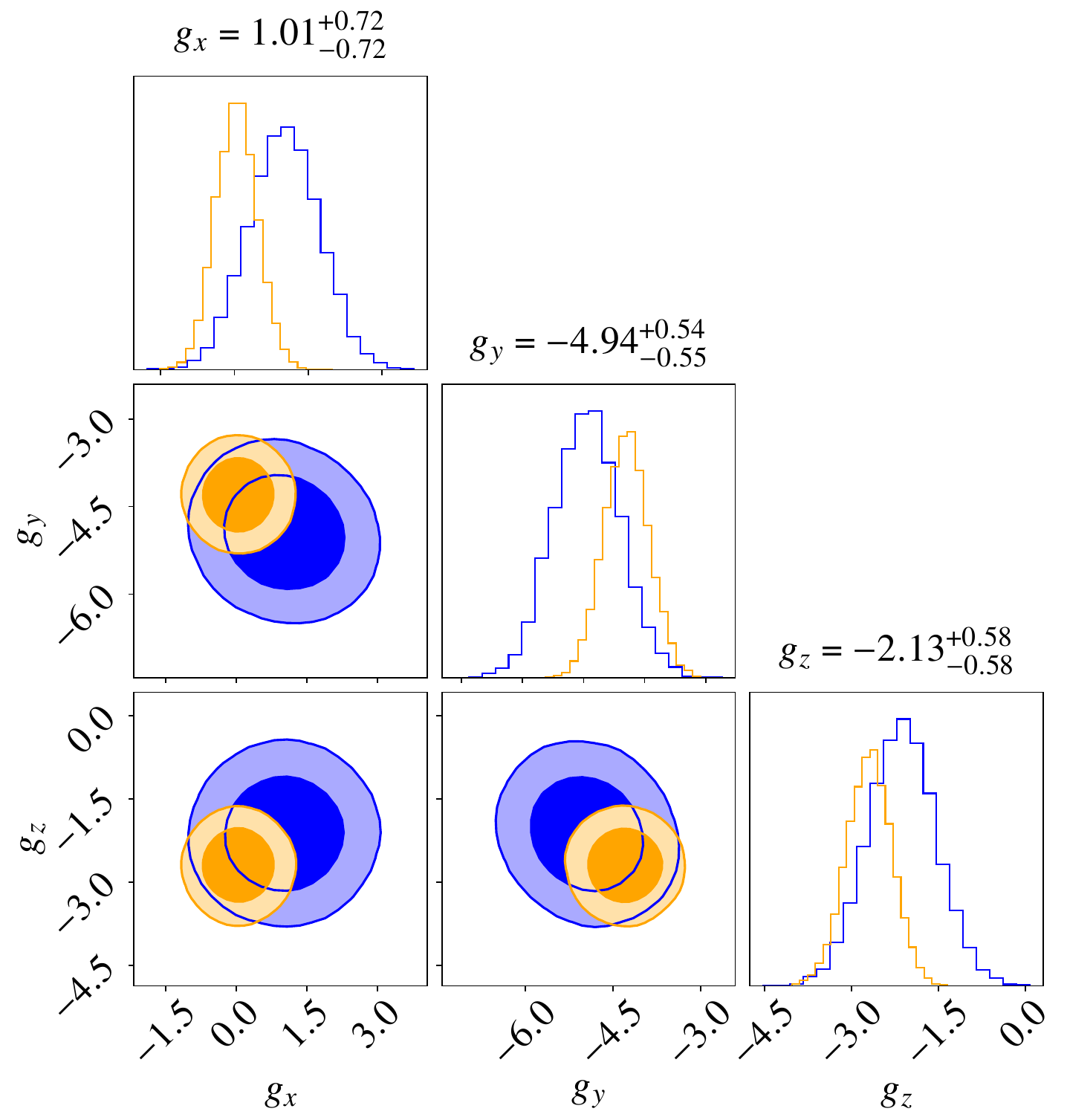}
        \label{subfig:d}
    }\hfill
    \subfloat[Quaia + $20^\circ$ galactic mask]{
        \includegraphics[width=0.32\linewidth]{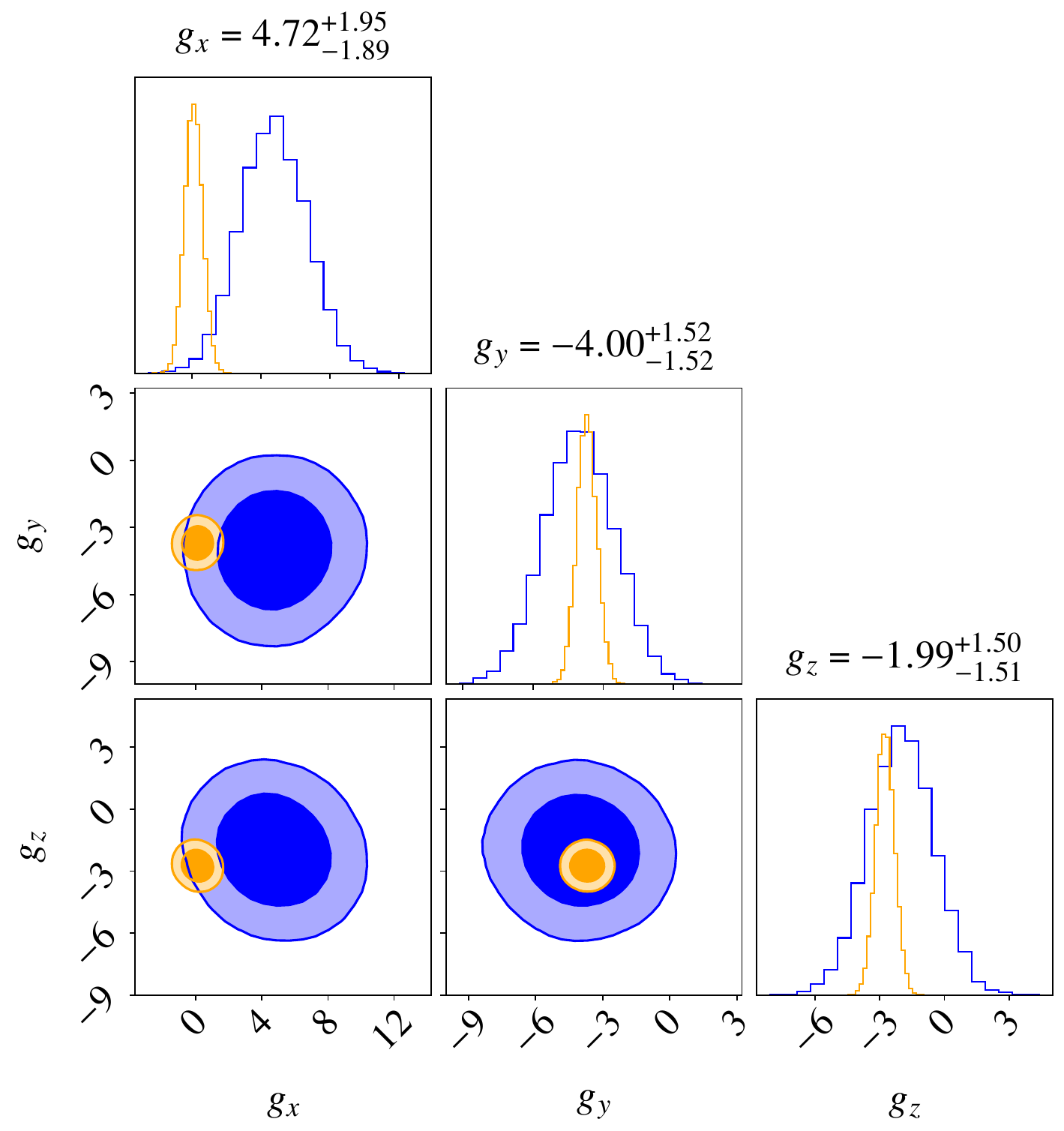}
        \label{subfig:e}
    }\hfill
    \subfloat[Quaia + $30^\circ$ galactic mask]{
        \includegraphics[width=0.32\linewidth]{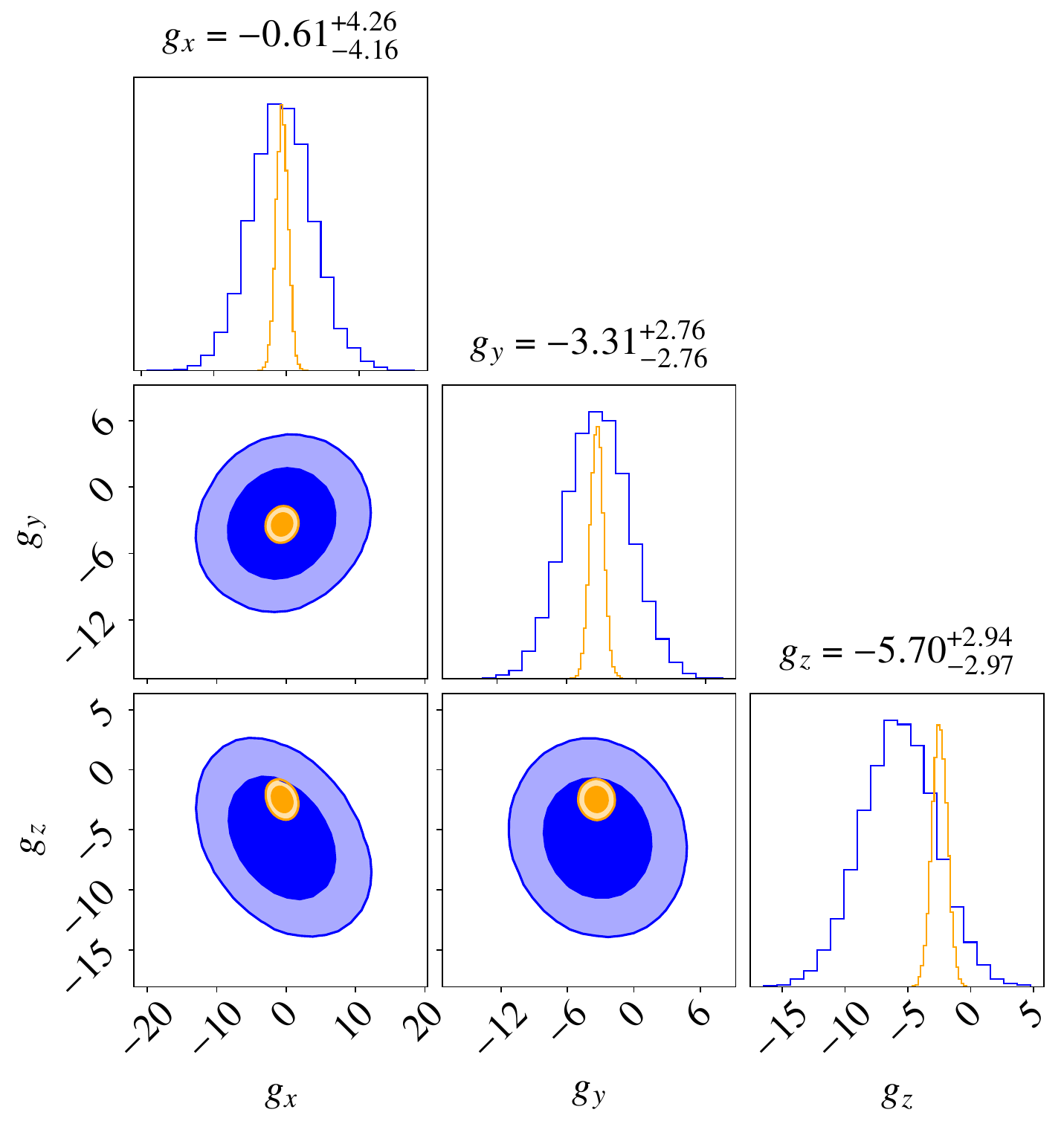}
        \label{subfig:f}
    }

    \caption{Posterior distributions for the acceleration components ($g_x, \,g_y,\,g_z$) from the SBI. The blue represents the input with VSH coefficients to $\ell = 3$ and $\hat{C}_\ell$ from $\ell = 4$. The orange represents input with $\ell_{max} = 3$ components. The contour marks 65 and 98 percentiles.}
    \label{fig:gxyz_corner}
\end{figure*}

One can see that when the extended fit is performed, the central values remain broadly stable (as shown in Fig.~\ref{fig:accel_compare_component}), but the credible intervals widen by factors of $\sim 1.5$ without masking, rising to $\sim 4$--$5$ under the $20^\circ$ and $30^\circ$ galactic cuts. The detection significance decreases accordingly, from $15.4\sigma$ to $12.4\sigma$ for unmasked Gaia EDR3, and from $13.7\sigma$ to $10.4\sigma$ for Quaia. The decrease is more pronounced under masking, dropping to $3.8\sigma$ ($4.1 \sigma$) and $3.2\sigma$ ($3.2 \sigma$) for the $20^\circ$ and $30^\circ$ galactic cut of Gaia EDR3 (Quaia). Our best estimate from Model 2 without additional galactic masking are ($g_x,\,g_y,\,g_z$) = $\left( 0.34^{+0.62}_{-0.59}, \,-4.52^{+0.46}_{-0.46},\,-2.59^{+0.43}_{-0.44} \right) \;\rm \mu as \,yr^{-1}$ for Gaia EDR3 and ($g_x,\,g_y,\,g_z$) = $\left(1.01^{+0.72}_{-0.72}, \,-4.94^{+0.54}_{-0.55},\,-2.13^{+0.58}_{-0.58}\right) \;\rm \mu as \,yr^{-1}$ for Quaia, with acceleration amplitude of $|\textbf{g}| = 5.28^{+0.43}_{-0.43}$ and $5.55^{+0.54}_{-0.53}\; \rm \mu as \,yr^{-1}$, respectively. The corresponding directions are $(\alpha,\, \delta) =
  \left(274.33^{+7.7}_{-7.5},\, -29.83^{+5.3}_{-4.8}\right)^\circ$ for Gaia
  EDR3 and $\left(281.39^{+7.8}_{-8.0},\, -22.93^{+6.5}_{-6.2}\right)^\circ$
  for Quaia, with $\sigma_{\rm GC} = 0.56 $ and $1.43$, respectively, both consistent with a Galactic-dominated acceleration. 

The most notable shift in Model 2 occurs for the $20^\circ$ galactic mask, where the $g_x$ for Gaia EDR3 (Quaia) shifts from $-0.08\,(0.11)$ to $4.07\, (4.72)$, which is a displacement of approximately $2.2 \sigma\,(2.4\sigma)$ relative to the Model 2 uncertainty. However, these shifts remain within the widened credible intervals and are therefore consistent with statistical noise from the reduced sky coverage and increased degeneracy between the dipole and higher-multipole nuisance parameters under aggressive masking.

\begin{table*}[t]
\caption{Solar System acceleration inferred from quasar proper motions.}
\label{tab:gxyz}
\centering
\renewcommand{\arraystretch}{1.55}
\resizebox{\linewidth}{!}{
\begin{tabular}{lccccccc}
\hline
\hline
  & \textbf{Quantity} &\textbf{Gaia EDR3} & \textbf{Gaia EDR3 + $20^\circ$ mask} & \textbf{Gaia EDR3 + $30^\circ$ mask} & \textbf{Quaia} &\textbf{Quaia + $20^\circ$ mask} & \textbf{Quaia + $30^\circ$ mask} \\
\hline
\multirow{10}{*}{\shortstack{\textbf{$\mathrm{a}_{\ell m}$}\\\textbf{$(\ell \leq 3)$}}}
 & $g_x$ & $-0.14_{-0.36}^{+0.36}$ & $-0.08_{-0.47}^{+0.47}$ & $-0.59_{-0.75}^{+0.77}$ & $0.04_{-0.42}^{+0.42}$ & $0.11_{-0.52}^{+0.52}$ & $-0.58_{-0.81}^{+0.83}$ \\
 & $g_y$ & $-4.06_{-0.31}^{+0.33}$ & $-3.45_{-0.38}^{+0.38}$ & $-3.02_{-0.57}^{+0.55}$ & $-4.30_{-0.37}^{+0.37}$ & $-3.70_{-0.41}^{+0.43}$ & $-3.37_{-0.59}^{+0.57}$ \\
 & $g_z$ & $-2.78_{-0.34}^{+0.34}$ & $-2.86_{-0.39}^{+0.38}$ & $-2.83_{-0.59}^{+0.58}$ & $-2.69_{-0.39}^{+0.38}$ & $-2.75_{-0.42}^{+0.44}$ & $-2.47_{-0.60}^{+0.62}$ \\
 & $|\mathbf{g}| $ & $4.94_{-0.31}^{+0.33}$ & $4.53_{-0.38}^{+0.38}$ & $4.30_{-0.54}^{+0.54}$ & $5.10_{-0.37}^{+0.38}$ & $4.66_{-0.42}^{+0.42}$ & $4.34_{-0.57}^{+0.58}$\\
 & $|\mathbf{g}|/\sigma_{|\textbf{g}|} $ & $15.4$ & $12.0$ & $7.9$ & $13.7$ & $11.0$ & $7.6$\\
 & $\alpha$ & $267.94_{-5.1}^{+5.1}$ & $268.66_{-7.6}^{+7.8}$ & $258.97_{-13.3}^{+14.5}$ & $270.55_{-5.6}^{+5.6}$ & $271.61_{-8.1}^{+8.2}$ & $260.36_{-12.7}^{+13.9}$\\
 & $\delta$ & $-34.36 _{-3.8}^{+4.0}$ & $-39.65_{-4.6}^{+5.1}$ & $-42.62_{-7.4}^{+8.9}$ & $-32.10_{-4.1}^{+4.5}$ & $-36.58_{-5.1}^{+5.6}$ & $-35.81_{-7.8}^{+9.1}$\\
 & $l$ & $356.02_{-4.0}^{+4.0}$ & $351.71_{-4.8}^{+4.7}$ & $345.26_{-7.2}^{+7.2}$ & $359.08_{-4.5}^{+4.5}$ & $355.55_{-5.4}^{+5.4}$ & $351.46_{-7.4}^{+7.5}$\\
 & $b$ & $-3.91_{-4.2}^{+4.1}$ & $-7.06_{-5.9}^{+6.1}$ & $-2.50_{-11.0}^{+10.9}$ & $-4.68_{-4.7}^{+4.7}$ & $-7.60_{-6.5}^{+6.6}$ & $0.54_{-11.5}^{+11.5}$\\
 & $\sigma_{\rm GC}$ & $0.89$ & $1.66$ & $1.58$ & $0.55$ & $0.96$ & $0.64$\\
\hline
\multirow{10}{*}{\shortstack{\textbf{$\mathrm{a}_{\ell m}\, (\ell \leq 3)$}\\\textbf{$+\,\hat{C}_\ell$}}}
 & $g_x$ & $0.34_{-0.59}^{+0.62}$ & $4.07_{-1.89}^{+1.83}$ & $-0.85_{-4.22}^{+4.24}$ & $1.01_{-0.72}^{+0.72}$ & $4.72_{-1.89}^{+1.95}$ & $-0.61_{-4.16}^{+4.26}$\\
 & $g_y$ & $-4.52_{-0.46}^{+0.46}$ & $-3.59_{-1.54}^{+1.49}$ & $-4.24_{-2.72}^{+2.83}$ & $-4.94_{-0.55}^{+0.54}$ & $-4.00_{-1.52}^{+1.52}$ & $-3.31_{-2.76}^{+2.76}$\\
 & $g_z$ & $-2.59_{-0.44}^{+0.43}$ & $-1.85_{-1.46}^{+1.43}$ & $-4.91_{-3.07}^{+2.96}$ & $-2.13_{-0.58}^{+0.58}$ & $-1.99_{-1.51}^{+1.50}$ & $-5.70_{-2.97}^{+2.94}$\\
 & $|\mathbf{g}| $ & $5.28_{-0.43}^{+0.43}$ & $6.14_{-1.59}^{+1.66}$ & $8.30_{-2.54}^{+2.73}$ & $5.55_{-0.53}^{+0.54}$ & $6.91_{-1.67}^{+1.69}$ & $8.35_{-2.54}^{+2.67}$ \\
 & $|\mathbf{g}|/\sigma_{|\textbf{g}|} $ & $12.4$ & $3.8$ & $3.2$ & $10.4$ & $4.1$ & $3.2$\\
 & $\alpha$ & $274.33_{-7.5}^{+7.7}$ & $317.33_{-18.3}^{+18.4}$ & $257.86_{-41.1}^{+61.8}$ & $281.39_{-8.0}^{+7.8}$ & $318.80_{-16.1}^{+16.3}$ & $256.82_{-48.0}^{+73.6}$\\
 & $\delta$ & $-29.83_{-4.8}^{+5.3}$ & $-18.96_{-13.2}^{+14.7}$ & $-48.72_{-12.3}^{+32.8}$ & $-22.93_{-6.2}^{+6.5}$ & $-17.86_{-11.9}^{+13.3}$ & $-59.08_{-7.6}^{+35.1}$\\
 & $l$ & $2.63_{-5.6}^{+5.7}$ & $29.63_{-19.2}^{+19.2}$ & $339.85_{-23.0}^{+23.3}$ & $11.71_{-6.8}^{+7.0}$ & $31.57_{-16.9}^{+17.4}$ & $330.99_{-23.7}^{+23.6}$\\
 & $b$ & $-6.44_{-6.1}^{+6.1}$ & $-38.65_{-15.1}^{+18.2}$ & $-5.42_{-34.3}^{+35.5}$ & $-9.01_{-6.9}^{+7.0}$ & $-39.57_{-13.8}^{+16.2}$ & $-10.98_{-32.4}^{+35.7}$\\
 & $\sigma_{\rm GC}$ & $0.56$ & $1.82$ & $0.42$ & $1.43$ & $2.16$ & $0.77$\\
\hline
\end{tabular}
}
\tablefoot{Acceleration components ($g_x$, $g_y$, $g_z$), magnitude $|\mathbf{g}|$ in $\rm \mu as\,yr^{-1}$, detection significance of the acceleration amplitude for each sample and masking configuration, and acceleration direction in equatorial $(\alpha,\,\delta)$ and Galactic $(l,\,b)$ coordinates (degrees). The quantity $\sigma_{\rm GC}$ denotes the tension between the recovered direction and the Galactic centre, expressed as an equivalent Gaussian significance. The upper block shows results from Model~1, and the lower block shows results from Model~2. Uncertainties are $1\sigma$ from the SBI posterior.}
\end{table*}

\section{Discussion}

\subsection{Comparing with previous determinations}

Our Model 1 fits reproduce the acceleration measurement of \citet{2021A&A...649A...9G}using bootstrap resampling for $\ell_{max} = 3$  across all samples and masking configurations. The central values of ($g_x, \, g_y,\, g_z$) are consistent within $1\sigma$. More broadly, they agree with the sequence of previous determinations, from the VLBI detections \citep{2011A&A...529A..91T, 2013A&A...559A..95T}  to the Gaia EDR3 measurement \citep{2021A&A...649A...9G}, as shown in Fig.~\ref{fig:accel_compare_component}. This confirms that the VSH decomposition and aberration drift signal are robust to the choice of quasar catalogue and sky coverage. The key differences lie in the uncertainty budget.

Bootstrap resampling draws individual sources with replacement from the observed catalogue and recomputes the VSH fit for each replicate, estimating the uncertainty from the spread of the resulting acceleration components across replicates. Consider the observed proper motion of source $i$:
\begin{equation}
\tilde{\boldsymbol{\mu}}_{i} = \boldsymbol{\mu}_{i} + \mathbf{n}_{s,i} + \mathbf{n}_{l,i}
\end{equation}
where $\boldsymbol{\mu}_{i}$ is the true proper motion signal, $ \mathbf{n}_{s,i}$ is the source-level measurement noise intrinsic to source $i$, and $\mathbf{n}_{l,i}$ is the large-scale correlated noise structure of the sky realisation. 

When bootstrap resamples individual sources, $\mathbf{n}_{s,i}$ varies between replicates since different sources are drawn each time. However, the large-scale noise ($\mathbf{n}_{l}$) is a property of the sky realisation rather than of individual sources. Since all bootstrap replicates are drawn from the same sky realisation, $\mathbf{n}_{l}$ is identical across all replicates and therefore has zero variance. This means that the large-scale noise does not contribute to the bootstrap uncertainty estimate, which captures only the source-level noise ($\mathbf{n}_{s}$) and the signal sampling variance. Bootstrap resampling therefore systematically underestimates the true uncertainty when spatially correlated noise is present

Therefore, when we extend the inference to jointly fit multipoles and marginalise over stochastic power $\hat{C}_\ell$ at $\ell = 4-10$, the credible intervals on the acceleration components widen by factors of 1.5 -- 5. This is the cost of accounting for the degeneracy between the dipole and higher-order structure in the proper motion field. We argue that the extended-fit uncertainties are more realistic than the bootstrap values. 

\subsection{Systematic templates} 
The cross-correlation analysis in Sec.~\ref {sec:cross-correlate} shows no statistically significant correlation at scale $\ell \geq 2$ between the quasar proper motion field and any of the three systematic templates across all samples and masking configurations. The $\rm M_{10}$ map and stellar density cross-correlation remain within $3 \sigma$ across all configurations, with most values well within $2\sigma$. The cross-correlation with the faint stellar proper motion map shows an isolated signal-to-noise excess at $\ell = 4$ for the unmasked and $20^\circ$ masked Gaia EDR3 and Quaia samples (see Fig.~\ref{fig:snr_per_ell} and Appendix~\ref{app:snr_cross_spectra}), which is suppressed within $3\sigma$ under $30^\circ$ galactic mask for both catalogues. However, the auto-power spectra in Fig.~\ref{fig:cl_auto} shows no corresponding excess at $\ell = 4$, indicating that this cross-correlation signal does not propagate into the quasar proper motion field itself and therefore does not affect the acceleration measurement. 

The null detection for the $\rm M_{10}$ map means that the scanning strategy does not produce a measurable imprint on the quasar proper motion field at any angular scale. This result contrasts with the suggestion of \citet{2022A&A...667A.148G} that the scanning law is responsible for the even-odd E-B asymmetry in the auto-power spectrum. 

Despite these null cross-correlations, the auto-power spectra in Fig.~\ref{fig:cl_auto}-\ref{fig:snr_auto_per_ell} show a marginal, systematically positive excess of power at $\ell \geq 2$ across all catalogue configurations and masking choices, which is not statistically significant (Appendix~\ref{app:snr_auto_spectra}). The three systematic templates tested do not account for this excess, and its origin remains unidentified. This leaves two possibilities open: either the signal originates from unidentified instrumental systematics not captured by the templates tested here, or it has a genuine astrophysical origin, such as large-scale anisotropies in the matter distribution, anisotropic Hubble expansion \citep{2019BAAS...51c.139D}, or a stochastic gravitational wave background (SGWB) contributing quadrupole power \citep{2011PhRvD..83b4024B} (discussed in Sec.~\ref{sec:GW}). Distinguishing between these scenarios requires both additional systematic templates and a physically motivated signal model for the higher multipoles, and we leave this to future work.

\subsection{Implications for the stochastic gravitational wave background} 
\label{sec:GW}

The quasar proper motion can, in principle, also carry the information about the SGWB at ultra-low frequencies. A passing gravitational wave induces correlated angular deflections in the positions of distant sources, and for SGWB, this mainly contributes to the quadrupole power in the proper motion field, with equal contribution from the E mode and B mode \citep{2011PhRvD..83b4024B, 2018ApJ...861..113D}.

\citet{2023MNRAS.524.3609J} use Gaia DR3 proper motions to constrain the SGWB by fitting a generic dipole + quadrupole pattern to several quasar datasets. For their cleanest dataset, they obtained a Z-score of 2.68 and Bayes factor of -23.5, which is consistent with noise. More recently, \citet{2025ApJ...982L..46D} improved this constraint by an order of magnitude by using the Quaia catalogue, obtaining $h_{70}^{2} \Omega_{GW} \leq 0.023$ (95\% confidence) from a quadrupolar VSH fit, and $h_{70}^{2} \Omega_{GW} \leq 0.0096$ from the first astrometric Hellings-Downs correlation analysis. This is roughly a factor of 4--9 below the \citet{2023MNRAS.524.3609J} limit. Neither analysis find any evidence for a SGWB signal. Fig.~\ref{fig:cl_auto} shows that the $\ell = 2$ relative E-mode and B-mode power depends on the masking configuration: for the unmasked Gaia EDR3 and Quaia samples, the E-mode exceeds the B-mode, while the masked samples show the opposite. This mask-dependent E/B ratio at $\ell =2$ is inconsistent with the fact that a SGWB is predicted to produce equal E mode and B mode power at $\ell =2$. This is consistent with the non-detection reported above.

\subsection{The acceleration as a function of redshift}

If the secular aberration drift is the sole origin of the observed proper motion dipole, the inferred acceleration must be independent of the redshift of the background sources, since the effect depends only on the observer's acceleration and the angular position of the source, not its distance. We tested this by dividing the Quaia sample into five redshift bins of equal source count without applying Galactic-plane cut and repeating the SBI analysis independently for each bin. Each bin contains about 126,880 sources, with bin edges $z = \{0.095, 0.915, 1.290, 1.638, 2.102, 4.435\}$.

\begin{figure}[tp!]
        \subfloat[Acceleration components]{
            \includegraphics[width=\linewidth]{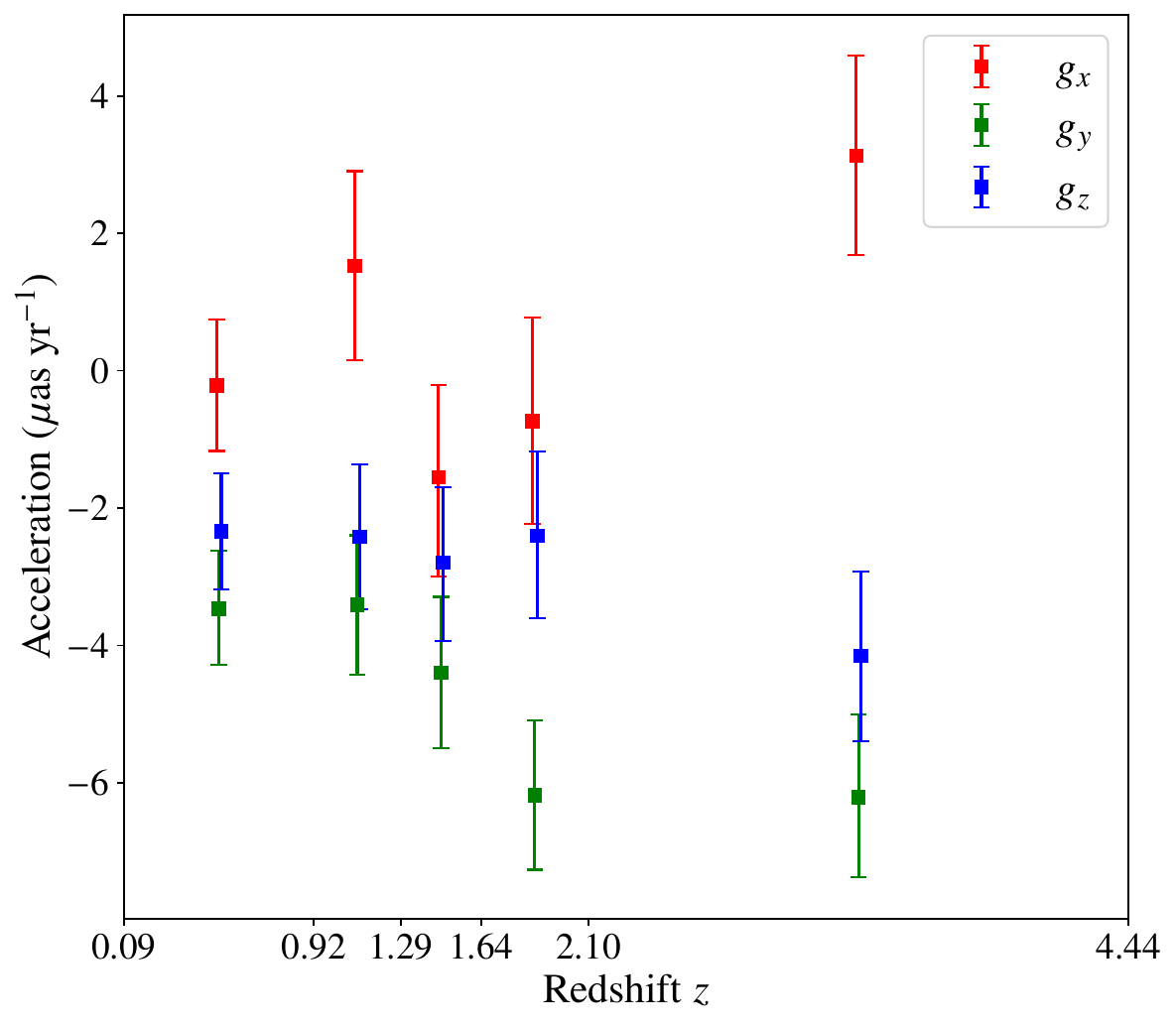}
            \label{subfig:accel_vs_z_a}
        }\vfill
        \subfloat[Acceleration amplitude]{
            \includegraphics[width=\linewidth]{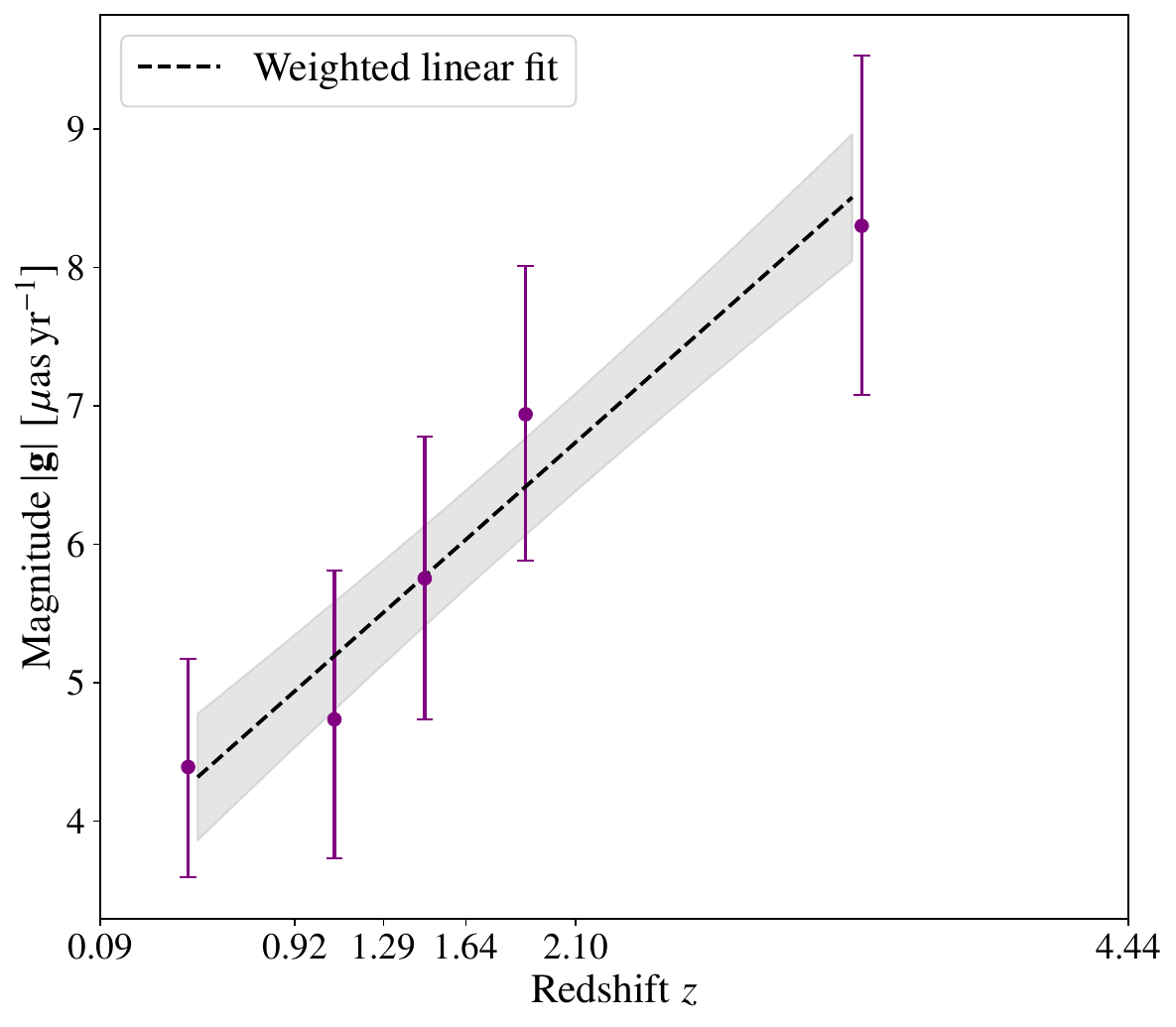}
            \label{subfig:accel_vs_z_b}
        }
        \caption{Posterior median values of acceleration (in $\mu\text{as},\text{yr}^{-1}$) derived from SBI applied to five equal-population redshift bins of the Quaia quasar catalogue. Error bars represent  
  the 16th--84th percentile credible interval of the marginalised posterior of the acceleration. \textit{Panel (a)}: Individual components $g_x$(red), $g_y$(green), and $g_z$(blue) with $\pm1\sigma$ error bars. \textit{Panel (b)}: Total acceleration amplitude $|g|$ with $\pm1 \sigma$ error bars. The black-dashed line represents the weighted linear fit line, with the shaded band indicating the $1\sigma$ propagated uncertainty on the fit. }
        \label{fig:accel_vs_z}
    \end{figure}

 The results are shown in Fig.~\ref{subfig:accel_vs_z_a}. To assess whether the inferred acceleration varies with redshift, we performed a $\chi^2$
   consistency test across the five redshift bins under the null hypothesis that the acceleration is redshift-independent.We find reduced   
  $\chi^2$ values of $1.78$, $1.80$, $0.43$, and $2.38$ for $g_x$, $g_y$, $g_z$, and $\mathbf{g}$, respectively, with the corresponding
 $p$-values of $0.13$, $0.13$, $0.79$, and $0.05$. All acceleration components are consistent with the null hypothesis at the $p > 0.05$ level. The      
  individual acceleration components ($g_x,\,g_y,\,g_z$) show no statistically significant dependence on
   redshift, in agreement with \citet{2011A&A...529A..91T}. However, the acceleration magnitude seems not to agree with it.
   
 As a complementary test on the total acceleration amplitude, we fitted a weighted linear model $|\mathbf{g}|(z) = mz  + c$ to the five redshift bins shown in Fig.~\ref{subfig:accel_vs_z_b}. The best-fit slope is $m = 1.52 \pm 0.51$, corresponding to $2.98\, \sigma$ departure from the null hypothesis $m = 0$. Although this falls short for a decisive detection, it still warrants discussion. We argue that the marginal positive slope of $2.98\, \sigma$ could reflect one of the following: (i) a statistical fluctuation consistent with noise, given the large uncertainties, particularly in the high-redshift bins; (ii) a residual systematic in the Quaia photometric redshift estimates, which carry a catastrophic outlier rate of $\sim 6 \%$ at $G < 20.0$ \citep{Storey_Fisher_2024} and could introduce a spurious redshift trend; or (iii) a genuine physical contribution beyond pure kinematic aberration, such as a redshift-dependent proper motion signal from large-scale structure. Such contributions are indeed present at some level: coherent bulk flows imprint correlated streaming proper motions on nearby sources, and the secular parallax induced by the solar motion relative to the cosmic microwave background produces a distance-dependent dipole of amplitude $78 \,\mu\mathrm{as\,yr}^{-1}\,(D/1\,\mathrm{Mpc})^{-1}$ \citep{2020ApJ...890..146P}. Both effects fall off with distance and are negligible at quasar distances ($\lesssim 0.05\,\mu\mathrm{as\,yr}^{-1}$ even at the median distance for our lowest-redshift bin). We do not interpret the marginal slope as evidence against a kinematic origin, but flag it as a result to be revisited with the spectroscopic quasar samples and improved proper motion baselines expected from Gaia DR4, which is anticipated for late 2026 and will provide 66 months of astrometric data \citep{2025arXiv250301533B} compared to the 34 months used here.

\section{Conclusions}

We have measured the acceleration of the Solar System with respect to the quasar rest frame using proper motions from Gaia, employing both the Gaia EDR3 quasar catalogue and the Quaia catalogue. Using the pseudo-$C_\ell$ framework, we have characterised the full multipole structure of the quasar proper motion field and its cross-correlations with the Gaia scanning strategy, faint stellar density and faint stellar proper motion. None of these templates produce a statistically significant detection in the signal-to-noise ratio analysis (all values are within $3\sigma$), yet the auto-power spectrum still shows marginal excess at $\ell \geq 2$ across all samples. This indicates that the origin of the higher-multipole power remains unidentified; in particular, it cannot be attributed to the Gaia scanning strategy, contrary to the suggestion of \cite{2022A&A...667A.148G} that the scanning law drives the even-odd E-B asymmetry.

By running two SBI pipelines - one with deterministic multipoles only (Model 1, $\ell_{max} =3$) and one that marginalises over stochastic higher-multipole power (Model 2, $\ell_{max} =10$) - we show that the acceleration amplitudes and directions are consistent with \citet{2021A&A...649A...9G} and with a Galactic centre-dominated acceleration, but the uncertainties widen by a factor of $\sim 1.5$ without masking, and by factors of $\sim 4$--$5$ under galactic cuts. For the Quaia catalogue, our Model 2 acceleration components are ($g_x,\, g_y, \,g_z$) = $\left( 1.01^{+0.72}_{-0.72},\,-4.94^{+0.54}_{-0.55},\,-2.13^{+0.58}_{-0.58}\right) \rm \;\mu as \,yr^{-1}$, directed towards $(\alpha,\, \delta) =
  (281.39^{+7.8}_{-8.0},\, -22.93^{+6.5}_{-6.2})^\circ$. We have also shown through a simple analytical argument that bootstrap resampling is insensitive to spatially correlated noise, explaining why previous uncertainty estimates were tighter.

The individual acceleration components show no statistically significant dependence on source redshift across five redshift bins, supporting the idea that the origin of the dipole is purely kinematic. A weighted linear fit to the total acceleration amplitude yields a marginal positive slope of $2.98 \sigma$, which we attribute to statistical noise and photometric redshift uncertainties in the Quaia catalogue rather than a genuine physical signal, though this warrants revisiting with future samples. 

The mask-dependent E/B power ratio at $\ell =2$ is inconsistent with the equal E- and B-mode power predicted by a SGWB, supporting the non-detection of SGWB reported by \citet{2023MNRAS.524.3609J} and \citet{2025ApJ...982L..46D}. Future Gaia data releases, with longer baselines and improved astrometric solutions, will further reduce the measurement uncertainties on the acceleration components, strengthen the redshift-independence test with more reliable redshift estimates, and may help identify the origin of the residual higher-multipole power.

\begin{acknowledgements}
We acknowledge APC for the funding at the beginning of the project. We also thank Ken Ganga, Alex Hall, Angelo Caravano, and Cotonou Alvarez Cardona for interesting discussions. This work relies on the open source packages \texttt{healpy} \citep{Zonca2019}, \texttt{numpy} \citep{2011CSE....13b..22V}, \texttt{sbi} \citep{Tejero-Cantero2020}, \texttt{Matplotlib} \citep{2007CSE.....9...90H}, \texttt{Scipy} \citep{2020NatMe..17..261V}, and \texttt{Namaster} \citep{Alonso_2019}.
\end{acknowledgements}

\bibliographystyle{aa}

\bibliography{ref}

\begin{appendix}

\section{Signal-to-noise ratio of the angular power spectra}

\subsection{Auto-power spectra} \label{app:snr_auto_spectra}

\begin{figure*}[htbp]
    \includegraphics[width=\linewidth]{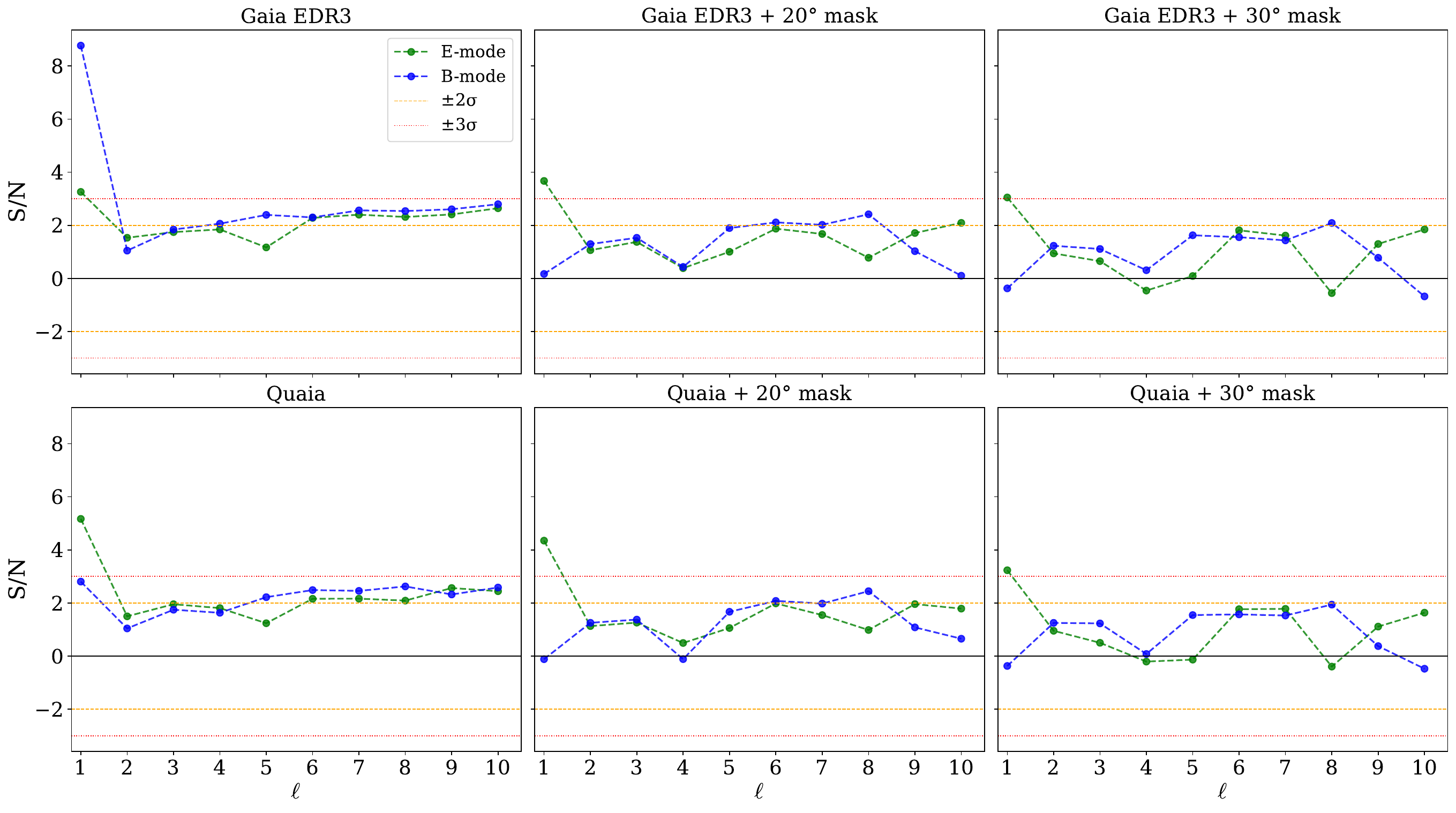}%
       
    \caption{Signal-to-noise ratio ($\hat{C}_\ell /\sigma_{\hat{C}_\ell}$) of the auto-power spectra. Green and blue dashed lines represent the E- and B-mode auto-power spectra of the proper motion map, respectively.}
    \label{fig:snr_auto_per_ell}
\end{figure*}

The signal-to-noise ratios of the auto-power spectra, defined as $\hat{C}_\ell/\sigma_{\hat{C}_\ell}$, are shown in Fig.~\ref{fig:snr_auto_per_ell} for all catalogue and masking configurations. The dipole $\ell = 1$ is detected at high significance, with the ratio exceeding $3 \sigma$ in all configurations, as expected from the Solar System's Galactic orbit. Several multipoles show a marginal positive excess over
  the noise-only expectation, but no single multipole exceeds $\sim
  3\sigma$; we therefore do not interpret the higher multipoles as an
  individually significant detection. However, we still accounted for it explicitly by marginalising over the stochastic power at $4 \leq \ell \leq 10$ in Model~2.

\subsection{Cross-power spectra} \label{app:snr_cross_spectra}
\begin{figure*}[htbp]
    \centering
    \includegraphics[width = \linewidth]{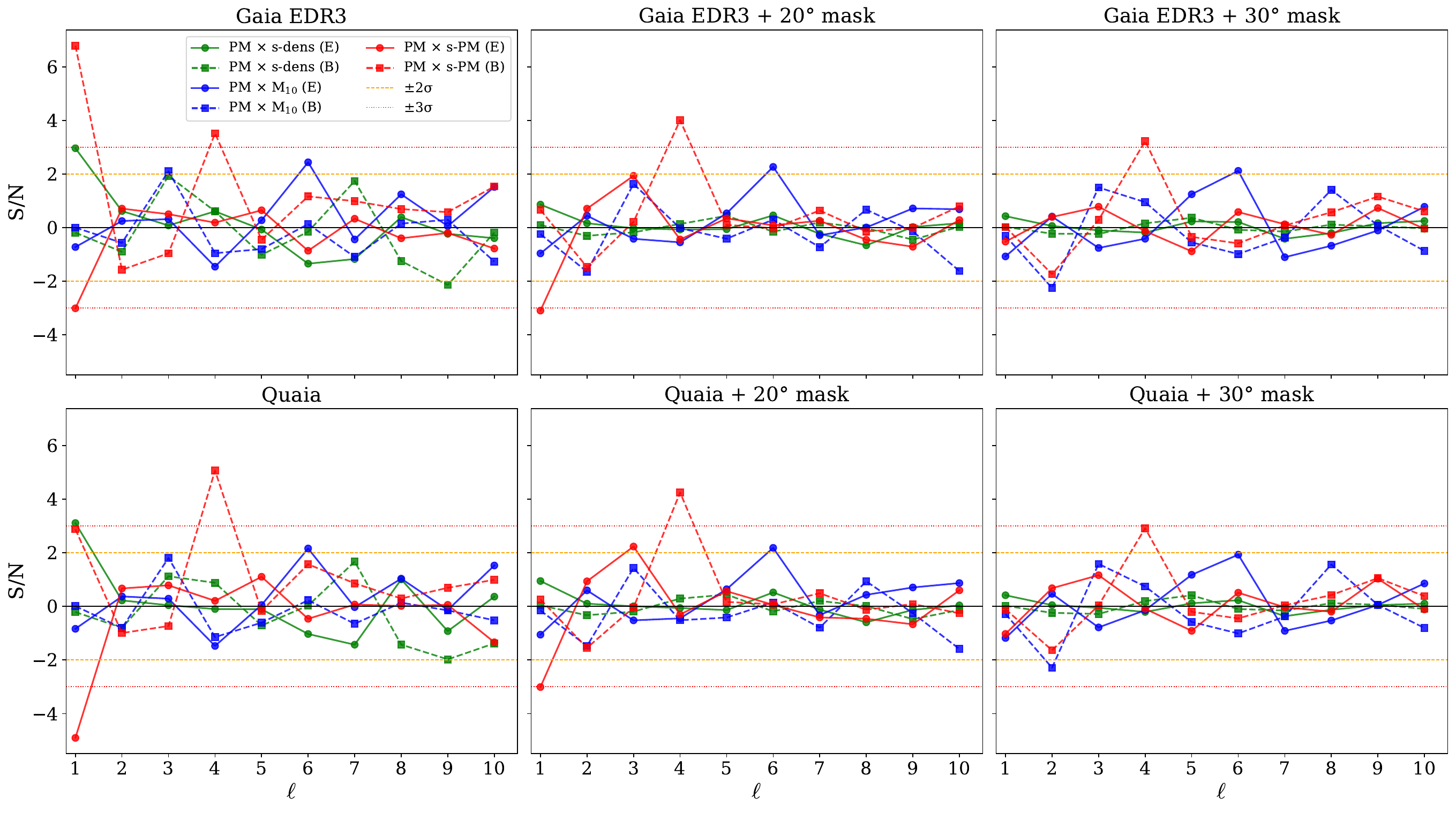}
    \caption{Signal-to-noise ratio ($\hat{C}_\ell /\sigma_{\hat{C}_\ell}$) of the cross-power spectra. Solid and dashed lines show the E mode and B mode, respectively. Green lines represent the cross-correlation with the faint stellar density map, red lines the cross-correlation with the faint stellar proper motion map, and blue lines the cross-correlation with the $\rm M_{10}$ map. Each panel corresponds to a different catalogue and masking configuration.}
    \label{fig:snr_per_ell}
\end{figure*}

The signal-to-noise ratios of the cross-power spectra between the quasar proper motion field and each systematic template are shown in Fig.~\ref{fig:snr_per_ell}. For the $\rm M_{10}$ scanning strategy map, the signal-to-noise ratio remains within $2\sigma$ across all multipoles and all catalogue and masking configurations, indicating no statistically significant correlation between the quasar proper motion field and the Gaia scanning strategy at any angular scale.

For the cross-correlation with the faint stellar density map, the signal-to-noise ratios at $\ell = 1$ reach marginally significant levels for the unmasked samples, approaching $3 \sigma$ at the dipole scale. Applying the galactic masks substantially suppresses the signal across all multipoles and configurations, bringing all values well within $2\sigma$. The cross-correlation with the faint stellar proper motion map shows an isolated excess of $\sim 4 - 5 \sigma$ at $\ell = 4$ for the unmasked and $20^\circ$ masked samples, while other multipole signals are well within $2\sigma$.

\section{Validation of the noise model and forward model} \label{app:validation}
~
\subsection{Pixel noise power spectrum} \label{app:pixel_noise}
\begin{figure*}[t!]
    \centering
    \includegraphics[width=\linewidth]{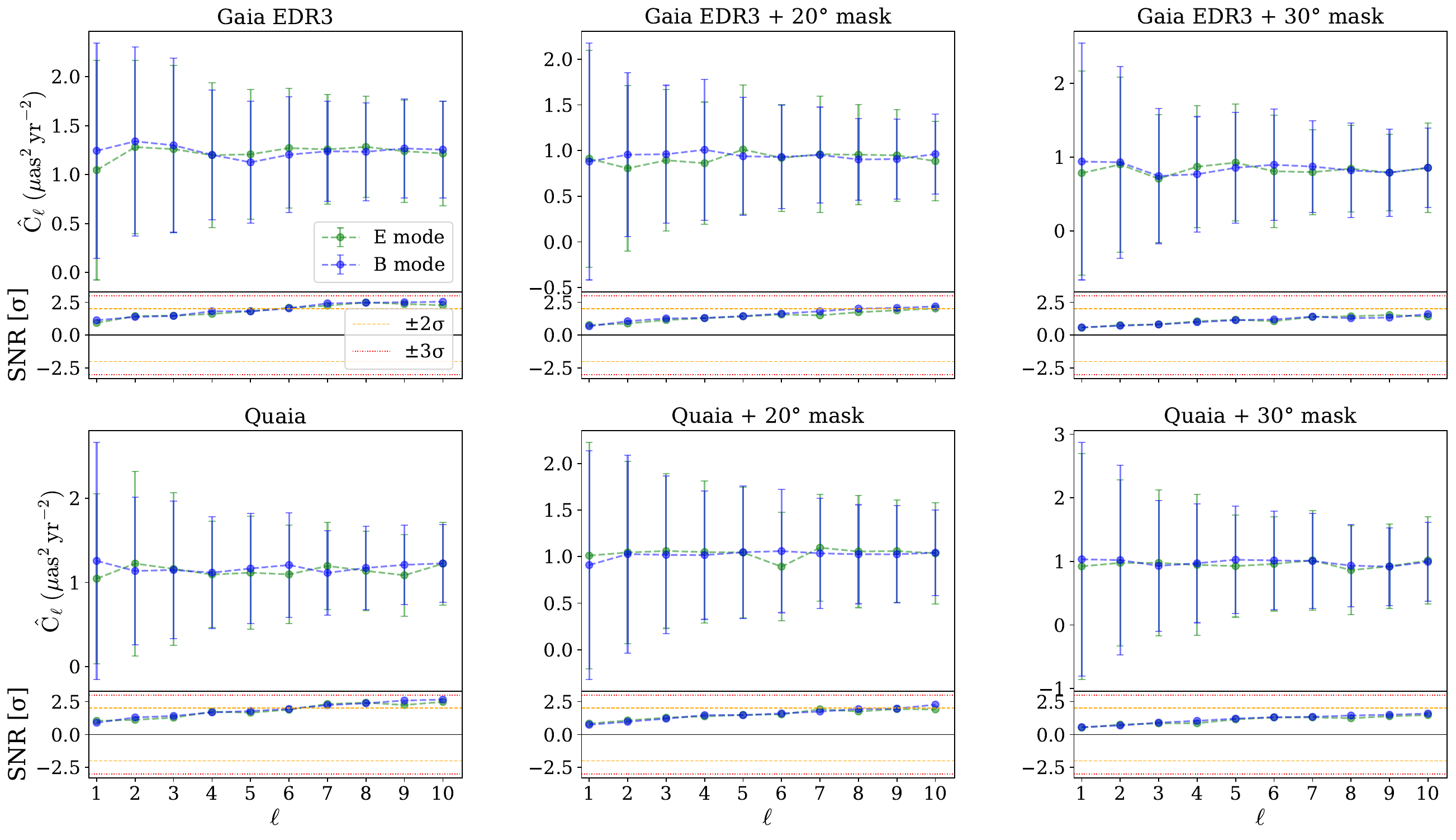}%

    \caption{Auto-power spectra (\textit{upper panels}) and signal-to-noise ratios (\textit{lower panels}) of the pixel noise contribution to the proper motion field, estimated from the per-pixel covariance matrices. E- and B-mode noise power are shown in blue and green, respectively.}
    \label{fig:pixel_noise_cl_and_snr}
\end{figure*}

~
To verify that the per-pixel noise model derived from the Gaia EDR3 covariance matrices is well behaved, we computed the auto-power spectrum of the map with only pixel noise. The resulting power spectra and their signal-to-noise ratios are shown in Fig.~\ref{fig:pixel_noise_cl_and_snr} for all data samples and masking configurations. The E-mode and B-mode power are consistent with each other and approximately scale independent across all multipoles. The signal-to-noise values remain within $ 3\sigma$, which confirms that the pixel noise is neither anomalously high nor spatially correlated at any angular scale. This validates the assumption made in the SBI that pixel noise can be modelled as the uncorrelated Gaussian contribution using the covariance matrices given by Eq.~\ref{eq:inverse_pixel_covariance}. 

\subsection{Forward model}  \label{app:pm_distribution}

\begin{figure*}
    \centering
    \includegraphics[width=1\linewidth]{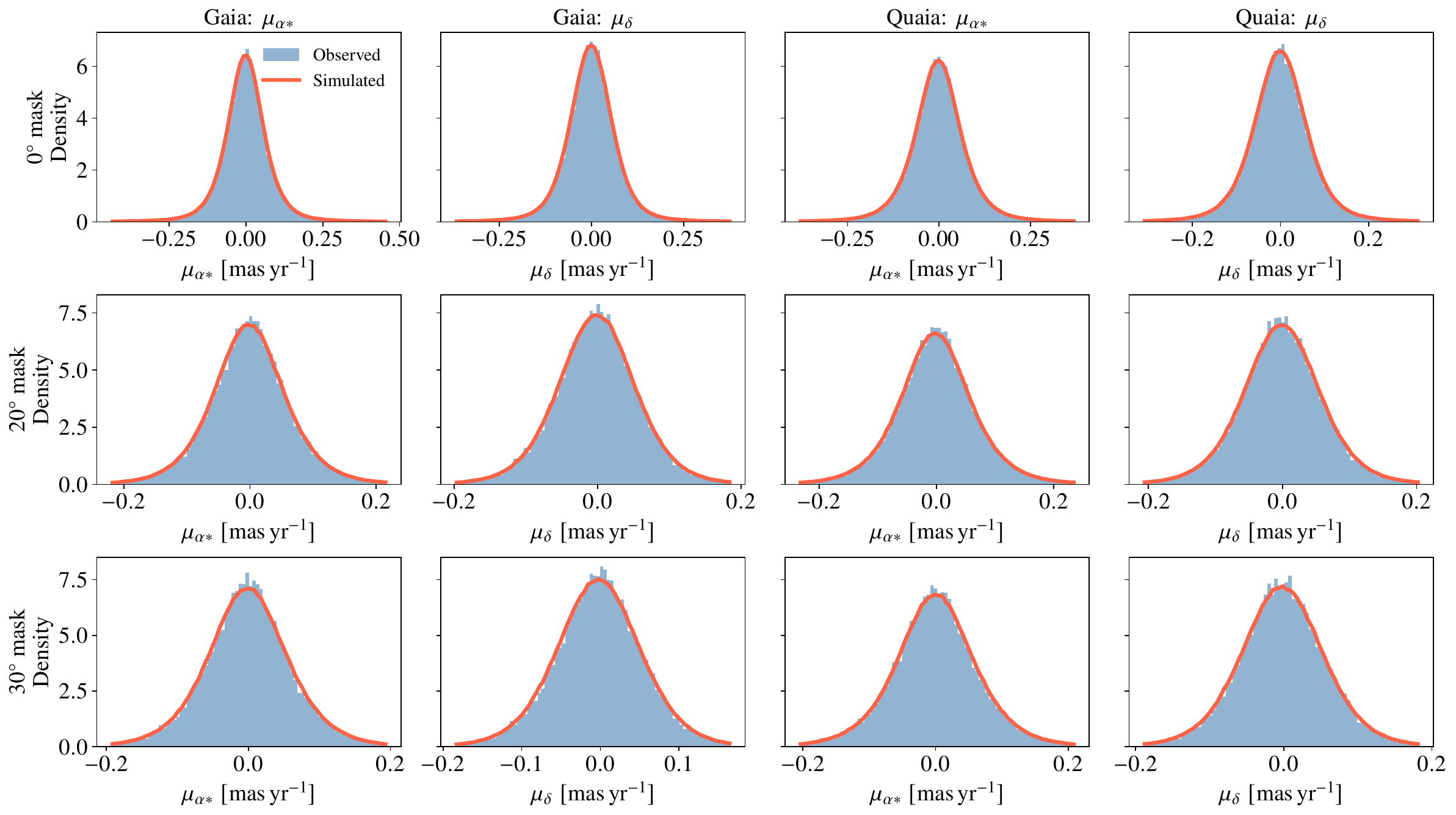}
    \caption{Distribution of the observed proper motion (blue) in right ascension ($\mu_{\alpha *}$) and declination ($\mu_{\delta}$), compared with the corresponding distributions predicted by the forward model (red). }
    \label{fig:pm_histogram}
\end{figure*}

We also compared the distributions of the observed quasar proper motions in right ascension and declination with the corresponding distributions predicted by the forward model. The result of the forward model is tested by calculating the best fit $a_{\ell m}$ using the solver in Sect.~\ref{sec:solver} up to $\ell_{max} =10$, then taking the coefficients to $\ell =3$ for the first part of the Model 2, then using $\hat{C}_{\ell}$ with $\ell = 4-10$ from Fig.~\ref{fig:cl_auto} for the second part. The comparison is shown in Fig.~\ref{fig:pm_histogram}. The forward model distributions closely match the observed distributions in both components across all samples, with no systematic offset or shape mismatch visible across all samples and masking configurations. This confirms that the noise model and the forward model pipeline provide an adequate statistical description of the data, and that the summary statistics fed into the NDE are not biased by a mis-specified model.

\section{Full posterior parameter space} \label{app:full_posterior}
\begin{figure*}
    \centering
    \includegraphics[width=\linewidth]{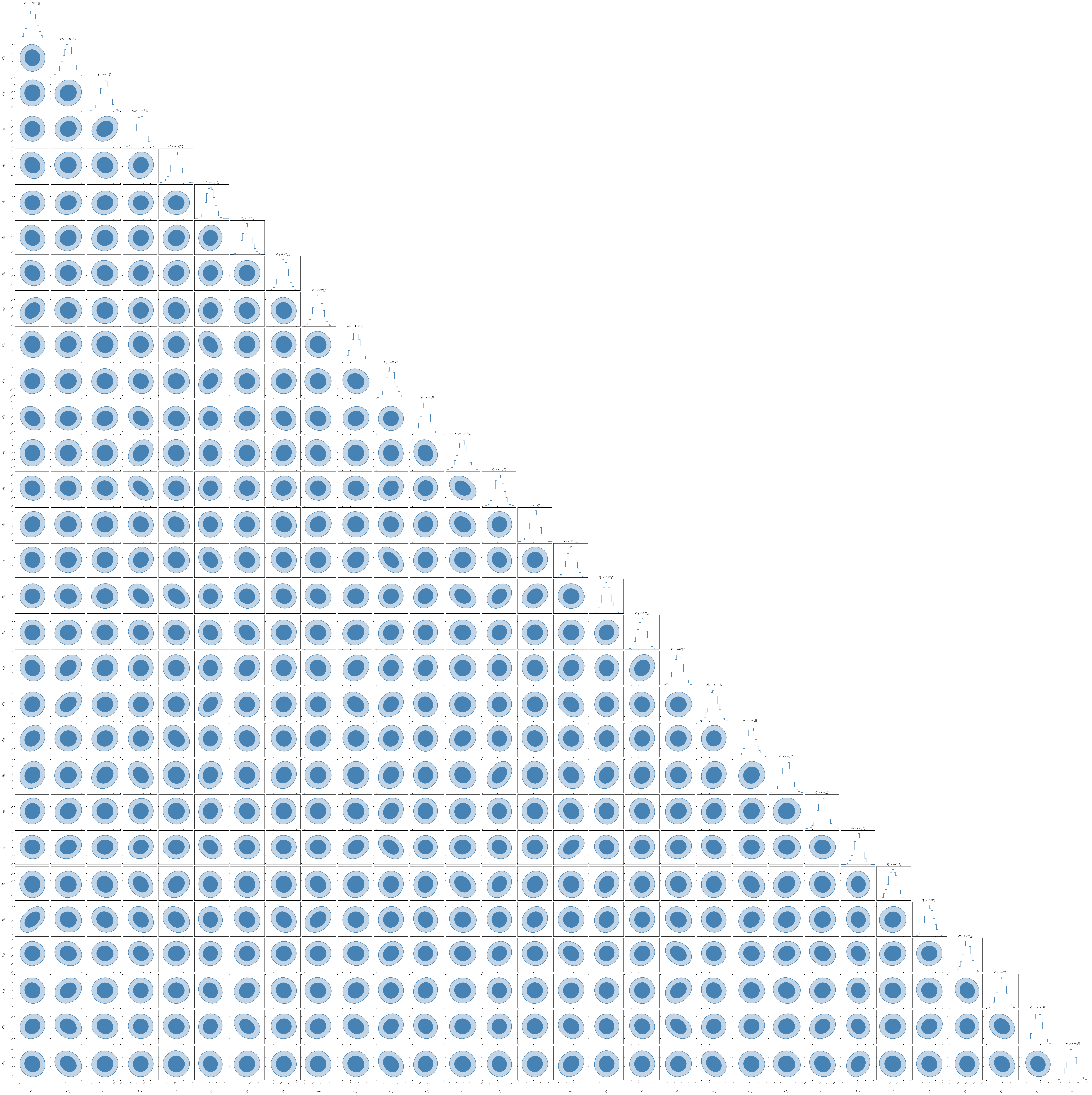}
    \caption{Posterior distribution of full $a_{\ell m}$ parameter space. The contours mark 65 and 98 percentiles.}
    \label{fig:corner_full_alm}
\end{figure*}

\begin{figure*}
    \centering
    \includegraphics[width=\linewidth]{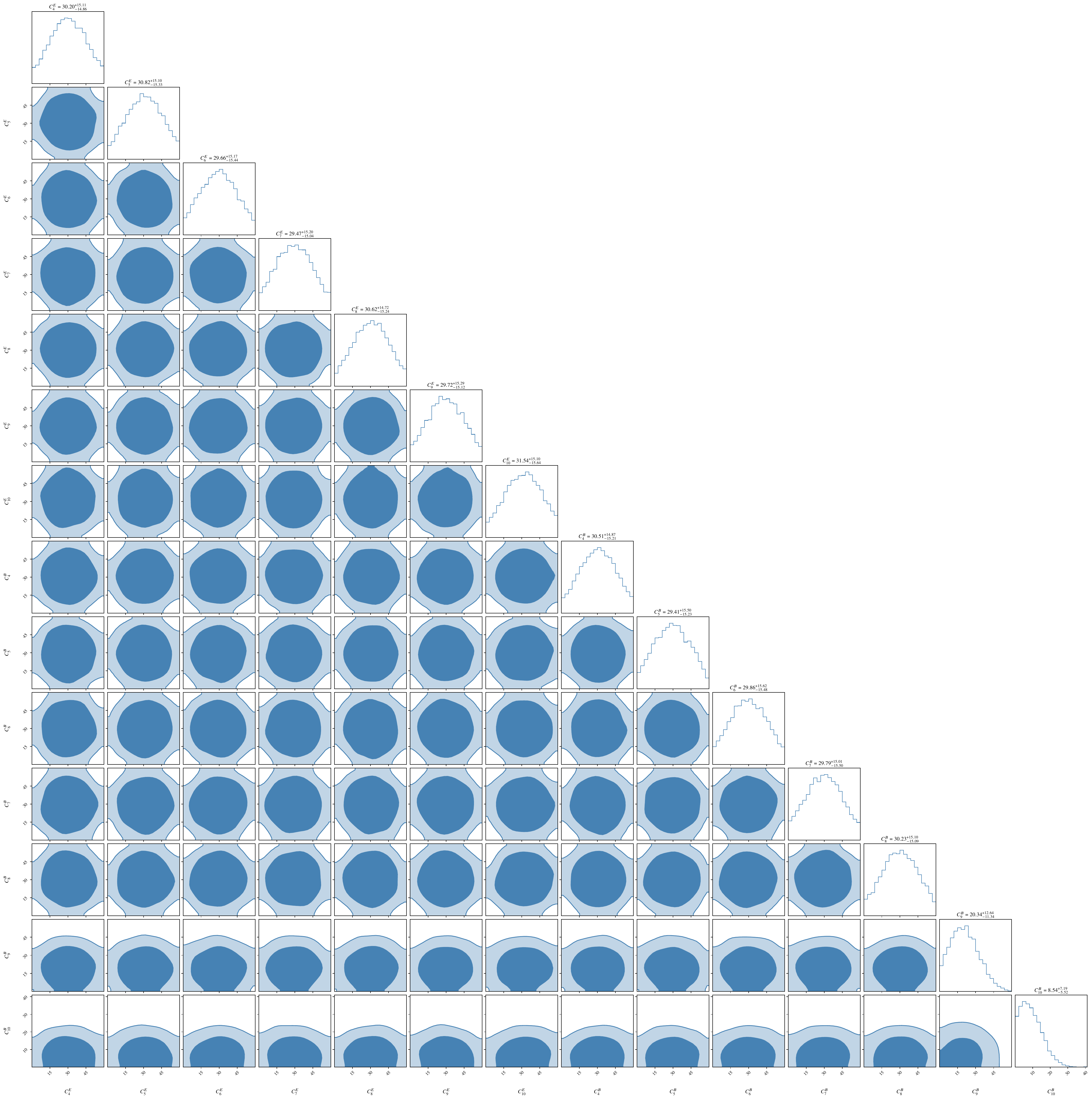}
    \caption{Posterior distribution of full $\hat{C}_{\ell}$ parameter space. The contour marks 65 and 98 percentiles.}
    \label{fig:cornet_full_cl}
\end{figure*}

The full posterior distributions of the Model 2 parameters inferred using the Quaia catalogue are shown in Fig.~\ref{fig:corner_full_alm} and Fig.~\ref{fig:cornet_full_cl}.

\end{appendix}

\end{document}